\documentclass[12pt]{iopart}

\usepackage{iopams}
\usepackage{graphicx}

\newcommand{\bmath}[1]{\mbox{{\boldmath{{$#1$}}}}}

\begin{document}

\title[Characterizing black holes with abundant gauge field hair]{Characterizing
asymptotically anti-de Sitter black holes
with abundant stable gauge field hair}

\author{Ben L. Shepherd and Elizabeth Winstanley}

\address{Consortium for Fundamental Physics, School of Mathematics and Statistics,
\\
The University of Sheffield,
Hicks Building, Hounsfield Road,
\\
Sheffield. S3 7RH United Kingdom}

\ead{E.Winstanley@sheffield.ac.uk}

\begin{abstract}
In the light of the ``no-hair'' conjecture, we revisit stable black holes in ${\mathfrak {su}}(N)$
Einstein-Yang-Mills theory with a negative cosmological constant $\Lambda $.  These black holes are endowed with
copious amounts of gauge field hair, and we address the question of whether these black holes can be uniquely characterized by
their mass and a set of global non-Abelian charges defined far from the black hole.
For the ${\mathfrak {su}}(3)$ case, we present numerical evidence that stable black hole configurations are fixed by their mass and
two non-Abelian charges.
For general $N$, we argue that the mass and $N-1$ non-Abelian charges are sufficient to characterize large stable black holes, in keeping with
the spirit of the ``no-hair'' conjecture, at least in the limit of very large $\left| \Lambda \right| $ and for a subspace containing stable black holes (and possibly some unstable ones as well).
\end{abstract}

\section{Introduction}
\label{sec:intro}

According to the ``no-hair'' conjecture \cite{Wheeler1}, equilibrium black holes
are extraordinarily simple objects, characterized completely by their mass,
angular momentum and charge.
These three quantities, as well as being parameters in the Kerr-Newman metric,
are also physical,
global charges, which could, at least in principle, be measured far
from the black hole.
For stationary, asymptotically flat black holes in four-dimensional Einstein-Maxwell
theory, the ``no-hair'' conjecture has been
proved (see for example \cite{no-hair} for reviews).
It is perhaps unsurprising that, if one of the above assumptions
(asymptotically flat space-time, four space-time dimensions,
Einstein-Maxwell theory)
is relaxed, then black hole uniqueness no longer holds.
Here we are particularly interested in matter theories other than Einstein-Maxwell,
and we focus on the
four-dimensional Einstein-Yang-Mills (EYM) model,
on which there is now an extensive literature (see,
for example, \cite{Volkov1,Winstanley3} for reviews).

The original discovery of soliton \cite{Bartnik1} and black hole \cite{Bizon1}
solutions of ${\mathfrak {su}}(2)$ EYM theory in
four-dimensional, asymptotically flat space, seemed, at least at first, to
provide counter-examples to the ``no-hair'' conjecture.
The black hole solutions have no global charges and are indistinguishable
from the usual Schwarzschild black holes at infinity.
However, both the black hole and soliton solutions are unstable \cite{Straumann1},
so while the ``letter'' of the ``no-hair'' conjecture (which,
in its original form, says nothing about stability) is violated, its ``spirit''
remains intact, because stable black holes do seem to be uniquely
characterized by global charges. These results led Bizon to reformulate the
``no-hair'' conjecture as follows \cite{Bizon2}:
\begin{quotation}
Within a given matter model, a {\em {stable}} stationary black hole is uniquely
determined by global charges.
\end{quotation}
For four-dimensional EYM theory in asymptotically flat space, this result
holds at least for spherically symmetric black holes
with a purely magnetic gauge field,
as the unique stable solution in this case is the Schwarzschild black hole{\footnote{It has been proven that all spherically symmetric, four-dimensional,
black hole solutions of EYM theory in asymptotically flat space with a
 purely magnetic gauge field
are unstable, for all gauge groups \cite{Brodbeck}.}}.
For asymptotically flat, four-dimensional EYM black holes with both an electric and magnetic field, the situation is less clear, with the
recent discovery \cite{Radu5} of stable dyonic black holes with a non-vanishing electric charge but zero magnetic charge in an extended EYM theory containing higher order terms in the field strength.

When a negative cosmological constant $\Lambda $ is introduced into the model,
so that the space-time is asymptotically anti-de Sitter (adS) rather than
asymptotically flat, stable soliton \cite{Bjoraker} and black hole solutions
\cite{Winstanley1} of ${\mathfrak {su}}(2)$ EYM theory exist.
For ${\mathfrak {su}}(2)$ gauge group, the purely magnetic gauge field is
described by a single function $\omega $.
Purely magnetic solutions have been proven to be stable \cite{Bjoraker,Winstanley1,Sarbach},
when the function $\omega $ has no zeros,
provided that $\left| \Lambda \right| $ is sufficiently large.
Unlike their asymptotically flat counterparts, these solutions do have a
global magnetic charge.

Enlarging the gauge group to ${\mathfrak {su}}(N)$, the purely magnetic
gauge field is now described by $N-1$ functions $\omega _{j}$.
For any fixed $N$, the existence of stable, purely magnetic,
soliton and black hole solutions
(for which all the $\omega _{j}$ have no zeros) has been proven
provided that $\left| \Lambda \right| $ is sufficiently large
\cite{Baxter1,Baxter2,Baxter3,Baxter4}.
These stable black holes have $N-1$ independent gauge degrees of freedom
and it turns out (see \cite{Baxter3} and section \ref{sec:numerical})
that they are described by $N+1$ independent parameters,
for gauge group ${\mathfrak {su}}(N)$.
We conclude \cite{Baxter1} that there is no limit to the amount of
stable gauge field hair with which a black hole in adS can be endowed.

Our purpose in this paper is to revisit these ${\mathfrak {su}}(N)$ EYM
black holes in the light of the ``no-hair'' conjecture as reformulated by Bizon.
In particular, we investigate whether these black holes are uniquely
characterized by global charges.
After a brief review of the ${\mathfrak {su}}(N)$ EYM model and its
black hole solutions in section \ref{sec:review}, we proceed, in section
\ref{sec:charges}, to construct global charges for the ${\mathfrak {su}}(N)$
gauge field.
We follow a two-pronged approach to examine whether these charges uniquely
characterize the black hole solutions, firstly, in section \ref{sec:numerical},
performing numerical investigations of the solution space, and then, in section
\ref{sec:analytic}, giving an analytic argument that
the black holes are uniquely characterized by global charges, at least for
sufficiently large $\left| \Lambda \right| $.
Our conclusions on the consequences of this result for the ``no-hair''
conjecture are presented in section \ref{sec:conc}.

\section{Hairy black holes in ${\mathfrak {su}}(N)$ Einstein-Yang-Mills theory
in anti-de Sitter space}
\label{sec:review}

In this section we outline the salient features of ${\mathfrak {su}}(N)$
EYM theory in anti-de Sitter space, and the black hole solutions
found numerically in \cite{Baxter2}, whose existence, for sufficiently large
$\left| \Lambda \right| $, was proven in \cite{Baxter3}.
We also discuss the thermodynamic properties of the black holes, and use the
boundary counter-term formalism \cite{Balasubramanian} to compute
their mass.

\subsection{Ansatz and field equations}
\label{sec:ansatz}

We consider four-dimensional ${\mathfrak {su}}(N)$ EYM theory
with a negative cosmological constant, described by the following action,
in suitable units:
\begin{equation}
S_{\mathrm {EYM}} = \frac {1}{2} \int d ^{4}x {\sqrt {-g}} \left[
R - 2\Lambda - {\mathrm {Tr}} \, F_{\mu \nu }F ^{\mu \nu }
\right] ,
\label{eq:action}
\end{equation}
where $R$ is the Ricci scalar, $\Lambda $ the cosmological
constant and ${\mathrm {Tr}}$ denotes a Lie algebra trace.
Throughout this paper, the metric has signature $\left( -, +, +, + \right) $ and
we use units in which $4\pi G = 1 = c$.
In addition, we fix the gauge coupling constant $g=1$.
In this article we focus on a negative cosmological constant, $\Lambda <0$.
Varying the action (\ref{eq:action}) gives the field equations
\begin{eqnarray}
T_{\mu \nu } & = & R_{\mu \nu } - \frac {1}{2} R g_{\mu \nu } +
\Lambda g_{\mu \nu };
\nonumber \\
0 & = & D_{\mu } F_{\nu }{}^{\mu } = \nabla _{\mu } F_{\nu }{}^{\mu }
+ \left[ A_{\mu }, F_{\nu }{}^{\mu } \right] ;
\label{eq:fieldeqns}
\end{eqnarray}
where the YM stress-energy tensor is
\begin{equation}
T_{\mu \nu } = F_{\mu \lambda }^{a} F_{\nu }^{a}{}^{\lambda }
- \frac {1}{4} g_{\mu \nu }
 F_{\lambda \sigma}^{a} F^{a\, \lambda \sigma } ,
\label{eq:Tmunu}
\end{equation}
with summation over the Lie-algebra index $a$ understood, so that $T_{\mu \nu }$ involves a Lie-algebra trace.
The Yang-Mills gauge field $F_{\mu \nu }$ is given in terms of the gauge potential
$A_{\mu }$ by
\begin{equation}
F_{\mu \nu } = \partial _{\mu }A_{\nu } - \partial _{\nu }A_{\mu } +
\left[ A_{\mu },A_{\nu } \right] .
\end{equation}

In this paper we are interested in static, spherically symmetric black hole
solutions of the field equations (\ref{eq:fieldeqns}), and we write the metric
in standard
Schwarzschild-like co-ordinates as:
\begin{equation}
ds^{2} = - \mu S^{2} \, dt^{2} + \mu ^{-1} \, dr^{2} +
r^{2} \, d\theta ^{2} + r^{2} \sin ^{2} \theta \, d\phi ^{2} ,
\label{eq:metric}
\end{equation}
where the metric functions $\mu $ and $S$ depend on the radial co-ordinate
 $r$ only.
In the presence of a negative cosmological constant $\Lambda <0$,
it is convenient to write the metric function $\mu $ as
\begin{equation}
\mu (r) = 1 - \frac {2m(r)}{r} - \frac {\Lambda r^{2}}{3}.
\label{eq:mu}
\end{equation}
We emphasize that in this paper we are considering only spherically symmetric
black holes and not topological black holes
which have been found in the ${\mathfrak {su}}(2)$ case \cite{Radu1}.

With a suitable choice of gauge, we take the purely magnetic
${\mathfrak {su}}(N)$ gauge potential to have the form \cite{Kunzle1}
\begin{equation}
A = \frac {1}{2} \left( C - C^{H} \right) \, d\theta - \frac {i}{2}
\left[ \left(
C + C^{H} \right) \sin \theta + D \cos \theta \right] \, d\phi ,
\label{eq:gaugepotsimple}
\end{equation}
where $C$ and $D$ are $\left( N \times N \right) $ matrices and
$C^{H}$ is the Hermitian conjugate of $C$.
The constant matrix $D$ takes the form:
\begin{equation}
D=\mbox{Diag}\left(N-1,N-3,\ldots,-N+3,-N+1\right) ,
\label{eq:matrixD}
\end{equation}
and the matrix $C$ is upper-triangular, with non-zero entries only
immediately above the diagonal:
\begin{equation}
C_{j,j+1}=\omega_j (r),
\label{eq:matrixC}
\end{equation}
for $j=1,\ldots,N-1$.
The gauge field is therefore described by the $N-1$ functions $\omega _{j}(r)$.
The derivation of the ansatz (\ref{eq:gaugepotsimple}) uses the Yang-Mills equations
and assumes that all the $\omega _{j}(r)$ are not identically zero
(see, for example, \cite{Galtsov1}
for the possibilities in asymptotically flat space if this
assumption does not hold).
We comment that our ansatz (\ref{eq:gaugepotsimple}) is by
no means the only possible choice in
${\mathfrak {su}}(N)$ EYM.
Techniques for finding {\em {all}} spherically symmetric ${\mathfrak {su}}(N)$
gauge potentials
can be found in \cite{Bartnik2}, where all irreducible models are
explicitly listed for $N\le 6$.

With the ansatz (\ref{eq:gaugepotsimple}),
there are $N-1$ non-trivial Yang-Mills equations
for the $N-1$ gauge field functions $\omega _{j}$:
\begin{equation}
r^2\mu\omega''_{j}+\left(2m-2r^3 p_{\theta}
-\frac{2\Lambda r^3}{3}\right)\omega'_{j}+W_j\omega_j=0
\label{eq:YMe}
\end{equation}
for $j=1,\ldots,N-1$, where a prime $'$ denotes $d /d r $,
\begin{eqnarray}
p_{\theta}&=&
\frac{1}{4r^4}\sum^N_{j=1}
\left[\left(\omega^2_j-\omega^2_{j-1}-N-1+2j\right)^2\right],
\label{eq:ptheta}
\\
W_j&=&
1-\omega^2_j+\frac{1}{2}\left(\omega^2_{j-1}+\omega^2_{j+1}\right),
\label{eq:Wdef}
\end{eqnarray}
and $\omega_0=\omega_N=0$.
The Einstein equations take the form
\begin{equation}
m' =
\mu G+r^2p_{\theta},
\qquad
\frac{S'}{S}=\frac{2G}{r},
\label{eq:Ee}
\end{equation}
where
\begin{equation}
G=\sum^{N-1}_{j=1}\omega_j'^2.
\label{eq:Gdef}
\end{equation}
The field equations (\ref{eq:YMe}, \ref{eq:Ee}) are invariant under the
transformation
\begin{equation}
\omega _{j} (r) \rightarrow -\omega _{j} (r)
\label{eq:omegaswap}
\end{equation}
for each $j$ independently, and also under the substitution:
\begin{equation}
j \rightarrow N - j.
\label{eq:Nswap}
\end{equation}

\subsection{Boundary conditions}
\label{sec:boundary}

Our primary interest in this paper is black hole solutions of the field equations
(\ref{eq:YMe}, \ref{eq:Ee}).
However, we will need to consider solitons in section \ref{sec:solitons} to ensure
that solitons cannot be mistaken for black holes by measuring global charges at
infinity.
The field equations are singular
at the origin, at an event horizon $r=r_{h}$ if there is one, and at infinity
$r\rightarrow \infty $.
Boundary conditions therefore have to be specified in a neighbourhood
of these singular points.
Local existence of solutions of the field equations satisfying the boundary
conditions outlined below is proven in \cite{Baxter3}.

\subsubsection{Origin}
\label{sec:bcorigin}

The boundary conditions at the origin are more complicated than near the event horizon
or at infinity. The full form is derived in detail in \cite{Baxter3}
(following the analysis of \cite{Kunzle2} for the asymptotically flat case).
Here we simply state the basic features which are needed for our analysis in section
\ref{sec:solitons}.
Near the origin, the field variables have the following form:
\begin{eqnarray}
m(r) & = & m_{3}r^{3} + O(r^{4});
\nonumber \\
S(r) & = & S_{0} + S_{2}r^{2} + O(r^{3});
\nonumber \\
\omega _{j}(r) & = & \pm \left[ j \left( N - j\right) \right] ^{\frac {1}{2}}
+ O(r^{2}).
\label{eq:origin1}
\end{eqnarray}
To fully specify the form of the gauge field in a neighbourhood of the origin,
a complicated power series has to be developed, up to $O(r^{N})$, the details
of which can be found in \cite{Baxter3} but which are not necessary for our purposes
in this paper.

\subsubsection{Event horizon}
\label{sec:bchorizon}

For black hole solutions, we assume that there is a regular, non-extremal
event horizon
at $r=r_{h}$, where $\mu (r)$ has a single zero.
This fixes the value of $m(r_{h})$ to be
\begin{equation}
m( r_{h} ) = \frac {r_{h}}{2} - \frac {\Lambda r_{h}^{3}}{6}.
\end{equation}
We assume that the field variables $\omega _{j}(r)$, $m(r)$ and $S(r)$
have regular Taylor series
expansions about $r=r_{h}$:
\begin{eqnarray}
m(r) & = & m (r_{h}) + m' (r_{h}) \left( r - r_{h} \right)
+ O \left( r- r_{h} \right) ^{2} ;
\nonumber \\
\omega _{j} (r) & = & \omega _{j}(r_{h}) + \omega _{j}' (r_{h})
\left( r - r_{h} \right)
+ O \left( r -r_{h} \right) ^{2};
\nonumber \\
S(r) & = & S(r_{h}) + S'(r_{h}) \left( r-r_{h} \right)
+ O\left( r - r_{h} \right) .
\label{eq:horizon}
\end{eqnarray}
Setting $\mu (r_{h})=0$ in the Yang-Mills equations (\ref{eq:YMe})
fixes the derivatives of the
gauge field functions at the horizon:
\begin{equation}
\omega _{j} ' (r_{h}) =
- \frac {W_{j}(r_{h})\omega _{j}(r_{h})}{2m(r_{h}) - 2r_{h}^{3}
p_{\theta } (r_{h})
-\frac {2\Lambda r_{h}^{3}}{3}}.
\label{eq:domegah}
\end{equation}
Therefore the expansions (\ref{eq:horizon}) are determined by
the $N+1$ quantities $\omega _{j}(r_{h})$, $r_{h}$, $S(r_{h})$ for fixed
cosmological constant $\Lambda $.
For the event horizon to be non-extremal, it must be the case that
\begin{equation}
2m'(r_{h}) = 2r_{h}^{2} p_{\theta} (r_{h}) < 1- \Lambda r_{h}^{2},
\label{eq:constraint}
\end{equation}
which weakly constrains the possible values
of the gauge field functions $\omega _{j}(r_{h}) $
at the event horizon.
Since the field equations (\ref{eq:YMe}, \ref{eq:Ee})
are invariant under the transformation
(\ref{eq:omegaswap}), we may consider
$\omega _{j}(r_{h}) >0$ without loss of generality.

\subsubsection{Infinity}
\label{sec:bcinfinity}

At infinity, we require that the metric (\ref{eq:metric}) approaches adS,
and therefore
the field variables $\omega _{j}(r)$, $m(r)$ and $S(r)$ converge to
constant values as
$r\rightarrow \infty $.
We assume that the field variables have regular Taylor series
expansions in $r^{-1}$ near infinity:
\begin{eqnarray}
m(r)  & = &  M + O \left( r^{-1} \right) ;
\qquad
S(r) = 1 + O\left( r^{-1} \right) ;
\nonumber \\
\omega _{j}(r) & = & \omega _{j,\infty } + c_{j}r^{-1} +
O \left( r^{-2} \right) .
\label{eq:infinity}
\end{eqnarray}
We have included the ${\cal {O}}(r^{-1})$ terms in $\omega _{j}(r)$
as they are central to our analysis in section \ref{sec:numerical}.
If the space-time is asymptotically flat, with $\Lambda =0$,
then the values of $\omega _{j,\infty }$
are constrained to be
\begin{equation}
\omega _{j,\infty } = \pm {\sqrt {j(N-j)}} .
\label{eq:AFinfinity}
\end{equation}
This condition means that the asymptotically flat black holes
have no magnetic charge at infinity (see section \ref{sec:charges} for the
definition of magnetic charges).
Therefore, at infinity, they are indistinguishable from
Schwarzschild black holes.
However, if the cosmological constant is negative,
then there are no {\it {a priori}} constraints on the values of
$\omega _{j,\infty }$.
In general, therefore, the adS black holes will be magnetically charged.
In section \ref{sec:charges} we will construct appropriate non-Abelian charges.

\subsection{Embedded solutions}
\label{sec:su2embedded}

The field equations (\ref{eq:YMe}, \ref{eq:Ee}) are non-linear and coupled,
but they do have two analytic, trivial solutions.
\begin{description}
\item[Schwarzschild-adS]
Setting
\begin{equation}
\omega _{j}(r) \equiv \pm {\sqrt {j(N-j)}}
\label{eq:SadS}
\end{equation}
for all $j$ gives the Schwarzschild-adS black hole
with
\begin{equation}
m(r) =M= {\mbox {constant.}}
\end{equation}
\item[Reissner-Nordstr\"om-adS]
Setting
\begin{equation}
\omega _{j}(r) \equiv 0
\label{eq:RNadS}
\end{equation}
for all $j$ gives the Reissner-Nordstr\"om-adS black hole
with metric function
\begin{equation}
\mu (r) = 1 - \frac {2M}{r} + \frac {Q^{2}}{r^{2}} -
\frac {\Lambda r^{2}}{3},
\end{equation}
where the magnetic charge $Q$ (see section \ref{sec:charges} for a definition of this quantity) is fixed by
\begin{equation}
Q^{2} = \frac {1}{6} N \left( N+1 \right) \left( N-1 \right) .
\end{equation}
Only for this value of the magnetic charge is the Reissner-Nordstr\"om-adS
black hole a solution
of the field equations.
\end{description}

As well as these effectively Abelian embedded solutions,
there is also a class of embedded ${\mathfrak {su}}(2)$
non-Abelian solutions, given by
writing the $N-1$ gauge field functions $\omega _{j}(r)$
in terms of a single function $\omega (r)$ as follows:
\begin{equation}
\omega _{j}(r) =\pm  {\sqrt {j(N-j)}} \, \omega (r)
\qquad \forall j=1,\ldots ,N-1.
\label{eq:embeddedsu2}
\end{equation}
It is shown in \cite{Baxter3} that, with a suitable rescaling of the other field
variables, the field equations (\ref{eq:YMe}, \ref{eq:Ee}) reduce to the
${\mathfrak {su}}(2)$ field equations for the function $\omega $.
Therefore any ${\mathfrak {su}}(2)$, asymptotically adS, EYM
black hole solution can be embedded
into ${\mathfrak {su}}(N)$ EYM to give an asymptotically adS black hole.

\subsection{Properties of the ${\mathfrak {su}}(N)$ solutions}
\label{sec:numerics}

The properties of soliton and black hole solutions of the field equations
(\ref{eq:YMe}, \ref{eq:Ee})
have already been studied in detail elsewhere
\cite{Winstanley3,Baxter1,Baxter2,Baxter3}, therefore here
we simply summarize the salient features required for our subsequent analysis.

For large $\left| \Lambda \right| $, numerical investigations
\cite{Baxter1,Baxter2} find both soliton and black hole
solutions for which all the gauge field functions $\omega _{j}(r)$ have no zeros.
It has been proven \cite{Baxter3} that, for fixed $r_{h}$
and $\omega _{j}(r_{h})$, black hole solutions for which
all the $\omega _{j}(r)$ have no zeros exist for all sufficiently
large $\left| \Lambda \right| $.
In view of these results, the focus in numerical work \cite{Baxter2}
has been on properties of the phase space of solutions
for fixed $r_{h}$ and different values of $\Lambda $.
In this work we are interested in the properties of the phase space
for fixed $\Lambda $ and varying horizon radius $r_{h}$.
We follow the standard method for finding numerical solutions (for further details, see \cite{Baxter2}).
Using a shooting method, the field equations (\ref{eq:YMe}, \ref{eq:Ee}) are integrated from close to the event horizon,
out towards infinity.

\begin{figure}
\begin{center}
\includegraphics[angle=270,width=8cm]{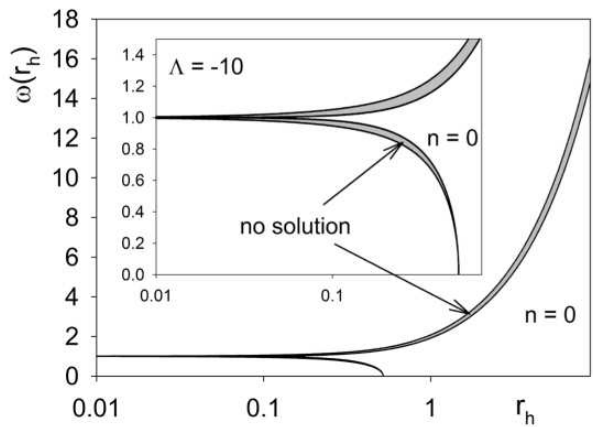}
\end{center}
\caption{Phase space of ${\mathfrak {su}}(2)$ solutions with
$\Lambda = -10$ and varying $r_{h}$.
The insert is a blow-up for smaller values of $r_{h}$ so that the
structure can be seen.   All the numerical solutions we find
are such that $\omega (r)$ has no zeros. The light grey region
corresponds to values of $\omega (r_{h})$ such that the condition
(\ref{eq:constraint}) is satisfied, but we do not find a numerical solution.
Below the lower boundary of
the upper grey region and above the upper boundary of the lower
grey region,
in the region labelled ``$n=0$'',
we find numerical solutions for which $\omega (r)$ has no zeros.
Above the upper grey region and below the lower grey region, the condition (\ref{eq:constraint})
for a regular event horizon is not satisfied.}
\label{fig:one}
\end{figure}

\begin{figure}
\begin{center}
\includegraphics[angle=270,width=8cm]{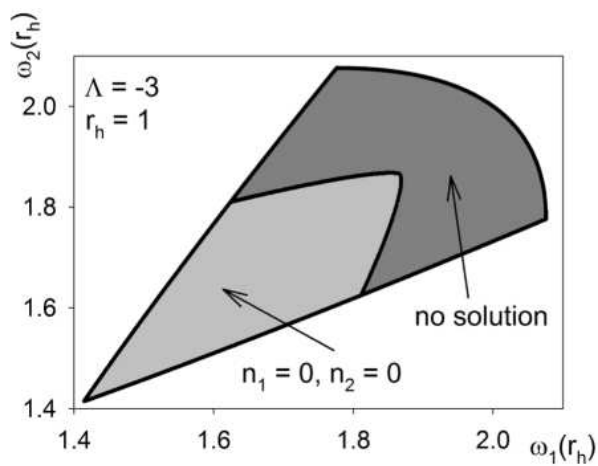}
\caption{Phase space of ${\mathfrak {su}}(3)$ solutions with $\Lambda = -3$
and $r_{h}=1$, with $n_{1}$ the number of zeros of the gauge field
function $\omega _{1}(r)$ and $n_{2}$ the number of zeros of $\omega _{2}(r)$.
We plot only the region in which the
inequalities (\ref{eq:stabineqs}) are satisfied on the event horizon.
The light grey region indicates solutions for which both the
gauge field functions $\omega _{1}(r)$ and $\omega _{2}(r)$ have no zeros.
The dark grey region corresponds to values of $\omega _{1}(r_{h})$
and $\omega _{2}(r_{h})$ for which the inequalities (\ref{eq:stabineqs})
 are satisfied on the event horizon, together with the condition (\ref{eq:constraint})
for a regular event horizon, but for which we do not find solutions.}
\label{fig:two}
\end{center}
\end{figure}

We find that, for sufficiently large $\left| \Lambda \right| $, all
numerical solutions for varying $r_{h}$ are such that the gauge
field functions $\omega _{j}(r)$ have no zeros.
For example, in figure \ref{fig:one} we show the phase space of
solutions for ${\mathfrak {su}}(2)$ black holes with $\Lambda = -10$ and
varying $r_{h}$.
Corresponding phase space studies for larger $N$ with $\Lambda $ fixed
and $r_{h}$ variable are more complex because of the number of parameters
involved, but in each case investigated we find similar results.
Phase space plots for $N>2$ can only be produced by fixing some of the
parameters, see, for example, figure \ref{fig:two}.

At least some of these black hole solutions for which all $\omega _{j}$
have no zeros are stable \cite{Baxter1,Baxter4}.
For ${\mathfrak {su}}(2)$, black hole solutions for which the single gauge
field function $\omega (r)$ has no zeros are shown to be stable if $\omega (r) > 1/{\sqrt {3}}$
everywhere on and outside the event horizon \cite{Winstanley1}.
For $N>2$, stability under linear, spherically symmetric perturbations
can be proven for sufficiently large
$\left| \Lambda \right| $, for ${\mathfrak {su}}(N)$ solutions
satisfying the inequalities \cite{Baxter1,Baxter4}:
\begin{equation}
\omega _{j}(r)^{2} > 1 + \frac {1}{2} \left[ \omega _{j+1}^{2}(r)
+ \omega _{j-1}^{2} (r) \right]
\label{eq:stabineqs}
\end{equation}
and which are sufficiently close to stable embedded ${\mathfrak {su}}(2)$
solutions.
The inequalities (\ref{eq:stabineqs}) have to be satisfied for all values
of $r\ge r_{h}$.
In figure \ref{fig:two} we plot the region of phase space in the
${\mathfrak {su}}(3)$
case, with $\Lambda = -3$ and $r_{h}=1$, for
which the inequalities (\ref{eq:stabineqs}) are satisfied on the black hole event
horizon and for which both $\omega _{1}(r)$ and
$\omega _{2}(r)$ have no zeros. For at least some of these solutions, the
inequalities (\ref{eq:stabineqs}) are satisfied for all $r\ge r_{h}$
\cite{Baxter1,Baxter4}.
However, the inequalities (\ref{eq:stabineqs}) are satisfied only in a
comparatively small region of the phase space for which there are nodeless
solutions (compare the size of the region in figure \ref{fig:two} with
figure 2 in \cite{Baxter3}, where the entire phase space for
${\mathfrak {su}}(3)$ black holes with $\Lambda = -3$ and $r_{h}=1$ is plotted).
Furthermore, due to the algebraic complexity of the gravitational sector
perturbation equations, stability for this sector can only be proven for
sufficiently large $\left| \Lambda \right| $ and for ${\mathfrak {su}}(N)$
solutions sufficiently close to stable embedded ${\mathfrak {su}}(2)$
solutions, with, as is usually the case in this kind of proof, little
indication of how large ``sufficiently large'' is, or how close
``sufficiently close''
is \cite{Baxter4}.
Our focus in this article is the characterization of stable ${\mathfrak {su}}(N)$
black holes. We therefore restrict attention to those black holes
for which the inequalities (\ref{eq:stabineqs}) are satisfied, at least on the
event horizon.  It is possible that our results include
some black holes which are in fact unstable.

\subsection{Thermodynamics of ${\mathfrak {su}}(N)$ EYM black holes}
\label{sec:thermodynamics}

\begin{figure}
\begin{center}
\includegraphics[width=6cm,angle=270]{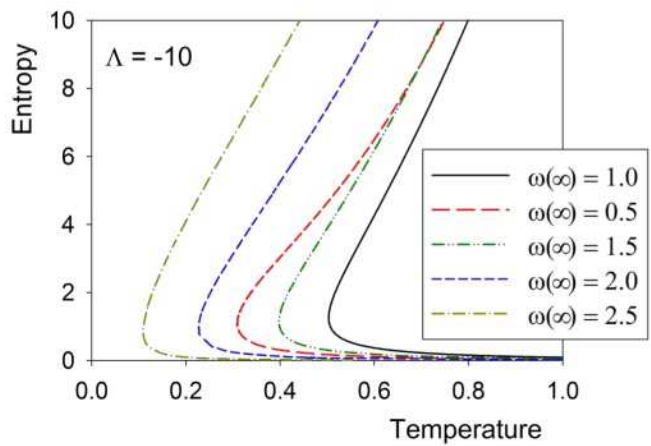}
\caption{Entropy-temperature curves for ${\mathfrak {su}}(2)$ black holes
with $\Lambda = -10$.
For each curve, the value of the gauge field function $\omega (r)$ at
infinity is fixed.
The curve with $\omega (\infty )=1$ corresponds to embedded
Schwarzschild-adS black holes.}
\label{fig:three}
\end{center}
\end{figure}

\begin{figure}
\begin{center}
\includegraphics[width=6cm,angle=270]{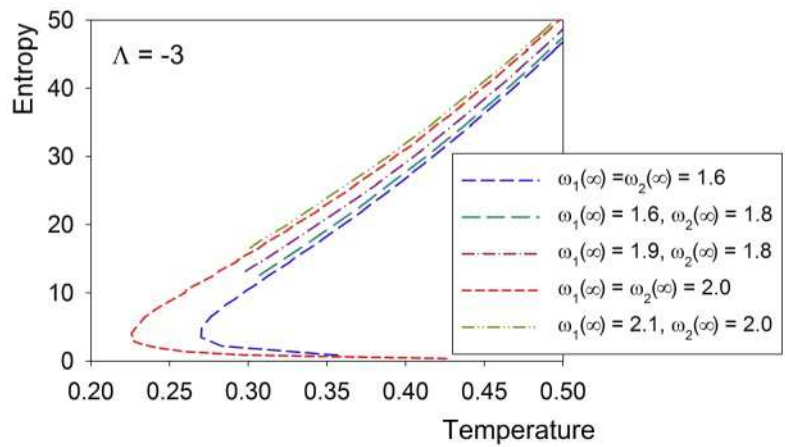}
\caption{Entropy-temperature curves for ${\mathfrak {su}}(3)$ black holes
with $\Lambda = -3$.
For each curve, the values of the gauge field functions $\omega _{1}(r)$
and $\omega _{2}(r)$ at infinity are fixed.
Curves with $\omega _{1}(\infty ) = \omega _{2}(\infty )$ correspond to
embedded ${\mathfrak {su}}(2)$ solutions.}
\label{fig:four}
\end{center}
\end{figure}

While the ``no-hair'' conjecture as outlined in the introduction is primarily
concerned with the characterization of black holes which are
{\em {classically}} stable, it nonetheless makes sense to consider the
thermodynamic stability of the black holes.
The entropy of ${\mathfrak {su}}(N)$ EYM black holes is given,
as usual, by one-quarter of the area of the event horizon, while their Hawking
temperature $T_{H}$ is
\begin{equation}
T_{H} = \frac {1}{4\pi r_{h}} \left( 1 - 2m'(r_{h}) - \Lambda r_{h}^{2} \right)
S(r_{h}),
\end{equation}
where we emphasize that $S(r_{h})$ is a metric function (\ref{eq:metric}),
and not the entropy of the  black hole.
For ${\mathfrak {su}}(2)$ black holes, the thermodynamics has already been
studied for $\Lambda = -3$ in \cite{Radu4}.
Their results are similar to ours in figure \ref{fig:three}
for the $\Lambda = -10$ case.
In figure \ref{fig:three}, we plot entropy as a function of
Hawking temperature for
fixed $\omega (\infty )$ (we will see in section \ref{sec:charges} that this
corresponds to fixing the magnetic charge of the black hole).
For each value of $\omega (\infty )$, there are two branches of
black hole solutions.
Firstly, there is a branch with small entropy and negative specific heat,
so that the black holes on this branch are thermodynamically unstable.
Secondly, there is an upper branch of black holes with larger entropy and
positive specific heat, corresponding to thermodynamically stable
black holes.
This is the behaviour found by \cite{Radu4} when $\Lambda = -3$, and
we found similar behaviour for other values of $\Lambda $.
The curve in figure \ref{fig:three} with $\omega (\infty ) =1$ corresponds
to embedded Schwarzschild-adS black holes.
As predicted in \cite{Visser} for generic hairy black holes in asymptotically
flat space, we see from figure \ref{fig:three} that the non-Abelian
${\mathfrak {su}}(2)$ black holes have lower temperatures than the embedded
Schwarzschild-adS black hole with the same entropy (and hence horizon area).
Similar behaviour is observed for ${\mathfrak {su}}(3)$ black holes, as can
be seen in figure \ref{fig:four}.
In figure 4, again anticipating the results of section \ref{sec:charges},
we fix the values of the two gauge field functions $\omega _{1}(r)$
and $\omega _{2}(r)$ at infinity, and then plot the curve of entropy as a
function of temperature.
If $\omega _{1}(\infty )= \omega _{2}(\infty )$, then we have embedded
${\mathfrak {su}}(2)$ solutions and the curves have a very similar
shape to those in figure \ref{fig:three}, with two branches of solutions.
In this case we are interested in the upper branch of solutions which are
thermodynamically stable.
For $\omega _{1}(\infty ) \neq \omega _{2}(\infty )$,
we have genuinely ${\mathfrak {su}}(3)$ solutions, and
in this case the curves appear to have just one branch of solutions,
which have positive specific heat and so are thermodynamically stable.
We are not able to say conclusively whether the fact that we have been
unable to find a thermodynamically unstable branch of solutions
is due to numerical difficulties or whether such a lower branch of
solutions does not in fact exist.

\subsection{Mass of ${\mathfrak {su}}(N)$ EYM black holes}
\label{sec:mass}

The mass of the non-Abelian black holes is readily computed using the
background counter-term formalism \cite{Balasubramanian}.
As is well known, the gravitational action for asymptotically adS space-times typically diverges as the boundary
$\partial {\mathcal {M}}$ of the
region (over which the Lagrangian is integrated) is taken to infinity.
This divergence is removed by the addition of boundary counter-terms to the
gravitational action, which do not alter the bulk equations of motion.
As we are working in four space-time dimensions,  the boundary counter-terms
which are sufficient to yield a finite bulk gravitational action
are \cite{Balasubramanian}:
\begin{equation}
I_{ct} = - \int _{\partial {\mathcal {M}}} d^{3} x \, {\sqrt {-\gamma }}
\left( \frac {2}{\ell } + \frac {\ell }{2} {\mathcal {R}} \right) ,
\label{eq:boundarycounterterms}
\end{equation}
where $\ell = {\sqrt {-\frac {\Lambda }{3}}}$ is the adS length
and the boundary metric $\gamma $ has Ricci scalar ${\mathcal {R}}$.
The boundary stress tensor, resulting from the variation of the total
gravitational action $S_{{\mathrm {grav}}}$ (which is the usual gravitational
action plus the counter-terms (\ref{eq:boundarycounterterms}))
is then  \cite{Balasubramanian}
\begin{equation}
T_{\mu \nu }^{B} = \frac {2}{{\sqrt {-\gamma }}}
\frac {\delta S_{{\mathrm {grav}}}}{\delta \gamma ^{\mu \nu }}
= \frac {1}{2} \left( \Theta _{\mu \nu } - \Theta \gamma _{\mu \nu }
 - \frac {2}{\ell } \gamma _{\mu \nu }
- \ell {\mathcal {G}}_{\mu \nu }
\right) ,
\end{equation}
where $\Theta _{\mu \nu }$ is the extrinsic curvature of the boundary
(with trace $\Theta $), and ${\mathcal {G}}_{\mu \nu }$ is the
Einstein tensor of the boundary metric.
Only the component $T_{tt}^{B}$ is required to compute the mass of the black holes.
To leading order, we find that
\begin{equation}
T_{tt}^{B} = \frac {M}{\ell r}
\end{equation}
where $M=\lim _{r\rightarrow \infty }m(r)$ (\ref{eq:infinity}), so that the
mass of the solutions is simply
\begin{equation}
\int _{\partial {\mathcal {M}}} \ell r T_{tt}^{B} \, d^{2}x = 4\pi M.
\end{equation}
From here on we will therefore use the variable $M$ to denote the ``mass'' of
the black holes.
Our results here are in complete agreement with those in \cite{Radu4}, where
it was found, for the ${\mathfrak {su}}(2)$ case,
that the non-Abelian gauge fields only contribute to the boundary stress
tensor at next-to-leading order, and therefore do not contribute
directly to the mass of the solutions.
Of course, they do contribute indirectly to the mass as they affect the
value of $M$ through the Einstein equations.

\section{Defining charges for ${\mathfrak {su}}(N)$ EYM black holes}
\label{sec:charges}

We now turn to the definition of global charges for ${\mathfrak {su}}(N)$
EYM black holes.
Since the rank of the ${\mathfrak {su}}(N)$ Lie algebra is $N-1$, we
expect these black holes to carry $N-1$ conserved charges,
which, at least in principle, could be measured at infinity, far from the
black hole.
The definition of charges for general non-Abelian gauge fields is
non-trivial because of the need for observable charges to be gauge-invariant,
coupled with the fact that the gauge field $F_{\mu \nu }$ is not itself
gauge-invariant in the non-Abelian case (unlike the
situation for Abelian gauge fields).
Methods for defining global, gauge-invariant, charges for non-Abelian gauge fields
have been devised by a number of authors \cite{charges,Brandt,Chrusciel,Lai}.
While the quantization of non-Abelian charge has been studied \cite{Brandt}, here we take a purely classical approach.

For an ${\mathfrak {su}}(2)$ gauge field, the single magnetic charge was defined in \cite{su2charges}
as
\begin{equation}
Q=\frac {1}{4\pi } \int _{S_{\infty }} {\sqrt {
F_{\theta \phi }^{a} F_{\theta \phi }^{a} }}
\, d\theta \, d\phi ,
\label{eq:su2charge1}
\end{equation}
where, under the square root, we have a Lie algebra trace over the $F_{\theta \phi }$ component of the gauge field, corresponding to a sum over the Lie algebra index $a$.
The integral is taken over the sphere at infinity.
The disadvantage of the formula (\ref{eq:su2charge1}) is that there is no natural generalization yielding $N-1$ charges in the ${\mathfrak {su}}(N)$
case.
An alternative ${\mathfrak {su}}(2)$ charge of the form
\begin{equation}
Q = 1 - \omega _{\infty }^{2}
\label{eq:su2charge2}
\end{equation}
has been considered by a number of authors (see, for example, \cite{Radu4}), and has the advantage of readily extending to the larger gauge group.
In particular, in this section we use the definitions of \cite{Chrusciel} (similar quantities were also defined in \cite{Lai})
to construct magnetic charges.
The approach of \cite{Chrusciel} has the advantage of yielding expressions which can be easily applied to our solutions.

Following \cite{Chrusciel}, we define gauge invariant magnetic charges as follows.
Let $X$ be an element in ${\mathfrak {A}}$, the Cartan sub-algebra of the
${\mathfrak {su}}(N)$ Lie algebra.
Then, for each $X$, a charge $Q(X)$ is defined by:
\begin{equation}
Q(X) = \frac{1}{4\pi} \sup_{g(r)} k\left(X, \int_{S_\infty} g^{-1}Fg \right)  .
\label{eq:qigen}
\end{equation}
Here the integral is taken over the sphere at infinity, and
the supremum is taken over all group elements $g(r)$ of the form
\begin{equation}
g(r) = \exp \left[ f(r) \Sigma \right]
\end{equation}
where $f(r)$ is a scalar function of $r$ and $\Sigma $ is a constant
element in the Lie algebra.
In (\ref{eq:qigen}), we have used $k(X,Y)=\Tr \left\{ {\mathrm {ad}} \,
X \, {\mathrm {ad}} \, Y \right\} $ which is the Killing form, with
${\mathrm {ad}} \, X$ denoting the adjoint representation of the Lie
algebra element $X$.
It is shown in \cite{Chrusciel} that the integrand in (\ref{eq:qigen})
 takes its maximal value when $g^{-1}Fg$ is in the Cartan sub-algebra
${\mathfrak {A}}$, and therefore we may restrict attention to those $g(r)$
for which this is the case.

Since we are integrating over a sphere at infinity, the integral in (\ref{eq:qigen})
simplifies to
\begin{equation}
Q(X)  =  \frac{1}{4\pi}k\left(X, \int_{S_\infty} g^{-1} F_{ \theta \phi} g \,
d\theta\, d\phi\right) .
\label{eq:qi}
\end{equation}
From the gauge potential ansatz (\ref{eq:gaugepotsimple}), the required
component of the field strength is found to be
\begin{equation}
F_{\theta \phi } = -\frac {i}{2} \left( \left[ C, C^{H} \right] - D \right)
\sin \theta ,
\label{eq:Fthetaphi}
\end{equation}
where $D$ is the constant matrix given in (\ref{eq:matrixD}).
From the form of the matrix $C$ (\ref{eq:matrixC}), it is straightforward
to show that
\begin{equation}
\left[ C,C^{H} \right] = {\mathrm {diag}}\left\{ \omega_1^2, \omega_2^2 -
\omega_1^2, \omega_3^2-\omega_2^2,\ldots ,-\omega_{N-1}^2 \right\}.
\end{equation}
We find that  $F_{\theta \phi }$ (\ref{eq:Fthetaphi}) is an element of the
Cartan sub-algebra ${\mathfrak {A}}$.
To see this, we begin by defining the following generators of the
Cartan sub-algebra \cite{Brandt}
\begin{equation}
H_{k} = -\frac {i}{{\sqrt {2k\left( k + 1 \right) }}} {\mathrm {diag}}
\left\{
{{\underbrace{1, 1, \ldots ,1}_{{k \, {\mathrm {entries}}}}}}, -k, 0, 0,
\ldots , 0
\right\} .
\end{equation}
Then $F_{\theta \phi }$ can be written in terms of the generators
$H_{k}$ as follows:
\begin{equation}
F_{\theta \phi } = \frac {\sin \theta }{2} \sum _{k=1}^{N-1} {\sqrt {2k
\left( k + 1 \right) }} \left( \frac {\omega _{k}^{2}}{k}
- \frac {\omega _{k+1}^{2}}{k+1} - 1\right) H_{k}.
\end{equation}
Since $F_{\theta \phi }$ is itself an element of the Cartan sub-algebra
${\mathfrak {A}}$, we can set $g(r)=e$, the identity element,
in (\ref{eq:qi}).
For each element of the Cartan sub-algebra $X\in {\mathfrak {A}}$,
equation (\ref{eq:qi}) yields a charge $Q(X)$, however only $N-1$ of these
will be independent because the rank of the ${\mathfrak {su}}(N)$
Lie algebra is $N-1$.
We therefore need a consistent way of selecting an appropriate basis of
charges $Q_{i}$, for $i=1,\ldots , N-1$.

First of all, we require our ${\mathfrak {su}}(N)$ charges (\ref{eq:qi}) to reduce
to (multiples of) the ${\mathfrak {su}}(2)$ charge (\ref{eq:su2charge2}) for embedded ${\mathfrak {su}}(2)$ solutions.
Secondly, we fix an overall normalization constant by defining a total
``effective'' charge $Q$ (reminiscent of the alternative ${\mathfrak {su}}(2)$ charge (\ref{eq:su2charge1})), as follows.
As $r\rightarrow \infty $, the metric function $m(r)$ takes the form
\begin{equation}
m(r) = M - \frac {Q^{2}}{2r} + O(r^{-2}) ,
\end{equation}
with $Q$ given by
\begin{equation}
Q^{2} = \sum _{i=1}^{N-1}Q_{i}^{2} = \frac {1}{2} \sum _{j=1}^{N} \left(
\omega _{j, \infty } ^{2} - \omega _{j-1, \infty }^{2}
-N - 1 +2j \right) ^{2} .
\label{eq:effectivecharge}
\end{equation}
Therefore $Q$ plays the same role in the metric as the usual Abelian charge.
For embedded Reissner-Nordstr\"om-adS black holes (\ref{eq:RNadS}), the expression (\ref{eq:effectivecharge}) reduces to
\begin{equation}
Q^{2} = \frac {1}{6}N \left( N - 1 \right) \left( N +1 \right) .
\end{equation}

With these two constraints, the natural basis of charges for ${\mathfrak {su}}(N)$ is
\begin{equation}
Q_{j} = \frac {{\sqrt {j\left( j + 1\right) }}}{{\sqrt {2}}} \left[ 1 -
\frac {\omega _{j,\infty } ^{2}}{j} +
\frac {\omega _{j+1 , \infty } ^{2}}{j+1} \right] ,
\label{eq:Qjfinal}
\end{equation}
for $j=1,\ldots , N-1$.
For the ${\mathfrak {su}}(2)$ case, there is just one charge
$Q_{1}=1-\omega (\infty )^{2}$, as required.
For the ${\mathfrak {su}}(N)$ case and embedded ${\mathfrak {su}}(2)$
black holes with $\omega _{j}(r) = {\sqrt {j\left( N - j\right)}} \omega (r)$,
the formula (\ref{eq:Qjfinal}) reduces to
\begin{equation}
Q_{j} = \frac {{\sqrt {j\left( j + 1 \right)}}}{{\sqrt {2}}} \left[ 1
 -\omega (\infty ) ^{2} \right]
\end{equation}
so that all the ${\mathfrak {su}}(N)$ charges are proportional to the
${\mathfrak {su}}(2)$ charge, again as required.
It is also straightforward to check that (\ref{eq:effectivecharge}) holds.

From (\ref{eq:Qjfinal}), it is clear that the charges are determined
uniquely by the values of the gauge field functions $\omega _{j}$ at
infinity. What is not so immediately apparent is that the converse is
 also true: the values of the gauge field functions at infinity can be determined
 (up to an overall irrelevant sign) from the charges $Q_{j}$ (\ref{eq:Qjfinal}).
In particular, from (\ref{eq:Qjfinal}), it can be shown that
\begin{equation}
\omega _{j,\infty }^{2} = j\left( N - j\right) -j {\sqrt {2}} \sum _{k=j}^{N-1}
 \frac {Q_{k}}{{\sqrt {k\left( k + 1 \right) }}} .
\label{eq:omegajcharge}
\end{equation}

\section{Characterizing ${\mathfrak {su}}(N)$ EYM black holes for large
$\left| \Lambda \right| $ - numerical work}
\label{sec:numerical}

Having defined a set of global charges for our ${\mathfrak {su}}(N)$ EYM black holes,
we now address the question of whether these charges,
together with the mass $M$ defined in section \ref{sec:mass} and the negative
cosmological constant $\Lambda $, are sufficient to characterize stable
${\mathfrak {su}}(N)$ EYM black holes.
We begin, in this section, with a numerical investigation, before turning,
in section \ref{sec:analytic}, to analytic arguments.

In \cite{Baxter3}, two local existence theorems are proved, for
solutions in a neighbourhood of the event horizon and infinity.
Near the event horizon, it is proven that the parameters $\Lambda $,
$r_{h}$ and $\omega _{j}(r_{h})$ (\ref{eq:horizon}) determine the solutions in a
neighbourhood
of the event horizon, and, furthermore, that these local solutions
are analytic in these parameters and the variable $r$.
Therefore $N+1$ parameters are required to completely specify the black hole
solutions\footnote{Strictly speaking, there is one further
parameter needed near the
horizon, namely $S(r_{h})$. However this parameter is fixed by the requirement
that $S\rightarrow 1$ as $r\rightarrow \infty $ and it plays no further role
in our analysis.}.

Near infinity, the situation is different.
It is proven in \cite{Baxter3} that $2N$ parameters are required to determine
the local solutions in a neighbourhood of infinity, namely $\Lambda $,
$M$, $\omega _{j,\infty }$ and $c_{j}$ (\ref{eq:infinity}).
Since we know that the black hole solutions form an $N+1$ parameter family,
it is clear that only $N+1$ of these $2N$ parameters are independent, but the
question is, which $N+1$ parameters can we take to be
a basis?
Ideally we would like the $N-1$ charges $Q_{j}$ (\ref{eq:Qjfinal}), defined in the previous section,
together with the mass $M$ and cosmological constant $\Lambda $, to be a
suitable basis.
From the analysis in the previous section, the charges $Q_{j}$ are completely
specified by the values of the gauge field functions
$\omega _{j,\infty }$ at infinity, so equivalently we would like the constants
$c_{j}$ in (\ref{eq:infinity}) to be determined
by $\Lambda $, $M$ and $\omega _{j,\infty }$.

\begin{figure}[h]
\begin{center}
\includegraphics[angle=270,width=12cm]{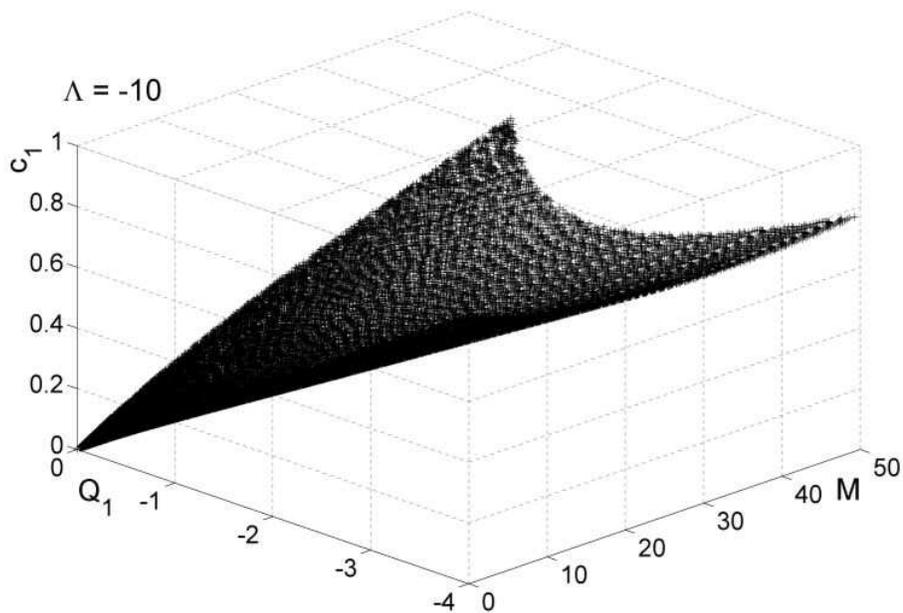}
\caption{$c_{1}$ (\ref{eq:infinity}) as a function of mass $M$ and charge $Q_{1}$ for stable ${\mathfrak {su}}(2)$ black hole
solutions with $\Lambda = -10$ and varying $r_{h}$. Only solutions for which the gauge field function $\omega (r)$
has no zeros and $\omega (r_{h}), \omega _{\infty }>1$ are plotted.  From this surface it is evident that $c_{1}$ can be regarded as a single-valued function of
$M$ and $Q_{1}$.}
\label{fig:su2c}
\end{center}
\end{figure}

\begin{figure}
\begin{center}
\includegraphics[angle=270,width=7cm]{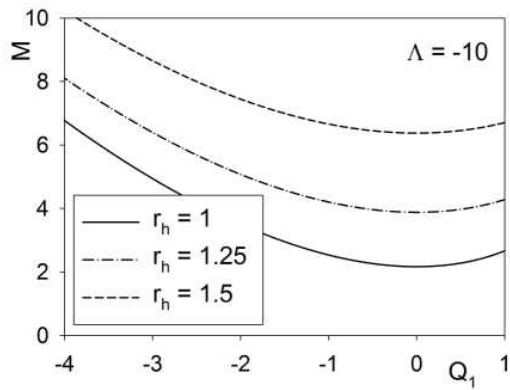}
\caption{Plot of $M$ against charge $Q_{1}$ for ${\mathfrak {su}}(2)$ black holes with $\Lambda = -10$ and varying $r_{h}$.
There is no evidence that the curves for different $r_{h}$ intersect, suggesting that $M$ and $Q_{1}$ uniquely
characterize the black holes.}
\label{fig:su2MQ}
\end{center}
\end{figure}

We begin our numerical investigations with the simplest case, namely ${\mathfrak {su}}(2)$ black holes.
In figure \ref{fig:su2c} we plot $c_{1}$ against the mass $M$ and charge $Q_{1}$ for stable ${\mathfrak {su}}(2)$ black hole
solutions when $\Lambda = -10$ (that is, we consider only solutions for which the gauge field function $\omega (r)$ has no zeros and
$\omega (r_{h}),\omega _{\infty }>1$).
It is clear from the surface in figure \ref{fig:su2c} that $c_{1}$ is a single-valued function of $M$ and $Q_{1}$.
We find similar results for other values of $\Lambda $.
For further evidence that $M$ and $Q_{1}$ therefore characterize ${\mathfrak {su}}(2)$ black holes, in figure \ref{fig:su2MQ}
we plot $M$ as a function of $Q_{1}$ for various $r_{h}$, with $\Lambda = -10$.
We have found no evidence that the $\left( Q_{1}, M \right)$ curves for different $r_{h}$ ever intersect. Therefore, for
each value of $\left( Q_{1}, M \right)$  for which there is a black hole solution, that solution is uniquely characterized
by $Q_{1}$ and $M$.

\begin{figure}[h]
\begin{center}
\includegraphics[angle=270,width=10cm]{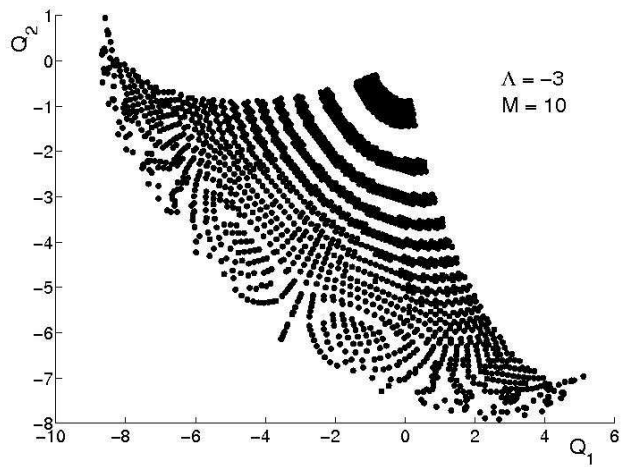}
\caption{Scatter plot of non-Abelian charges $Q_{1}$ and $Q_{2}$ for potentially stable ${\mathfrak {su}}(3)$ black holes with $\Lambda = -3$ and mass
$M=10\pm 0.1$. We perform a scan over the black hole solutions, and each data point represents a numerical black hole solution.
The discreteness of the grid we use can be seen as the values of the charges become more negative.
We find bands of charge values, although the bands seem to merge and the structure becomes less clear as the charges become more negative.}
\label{fig:su3q1q2}
\end{center}
\end{figure}

\begin{figure}
\begin{center}
\includegraphics[angle=270,width=7cm]{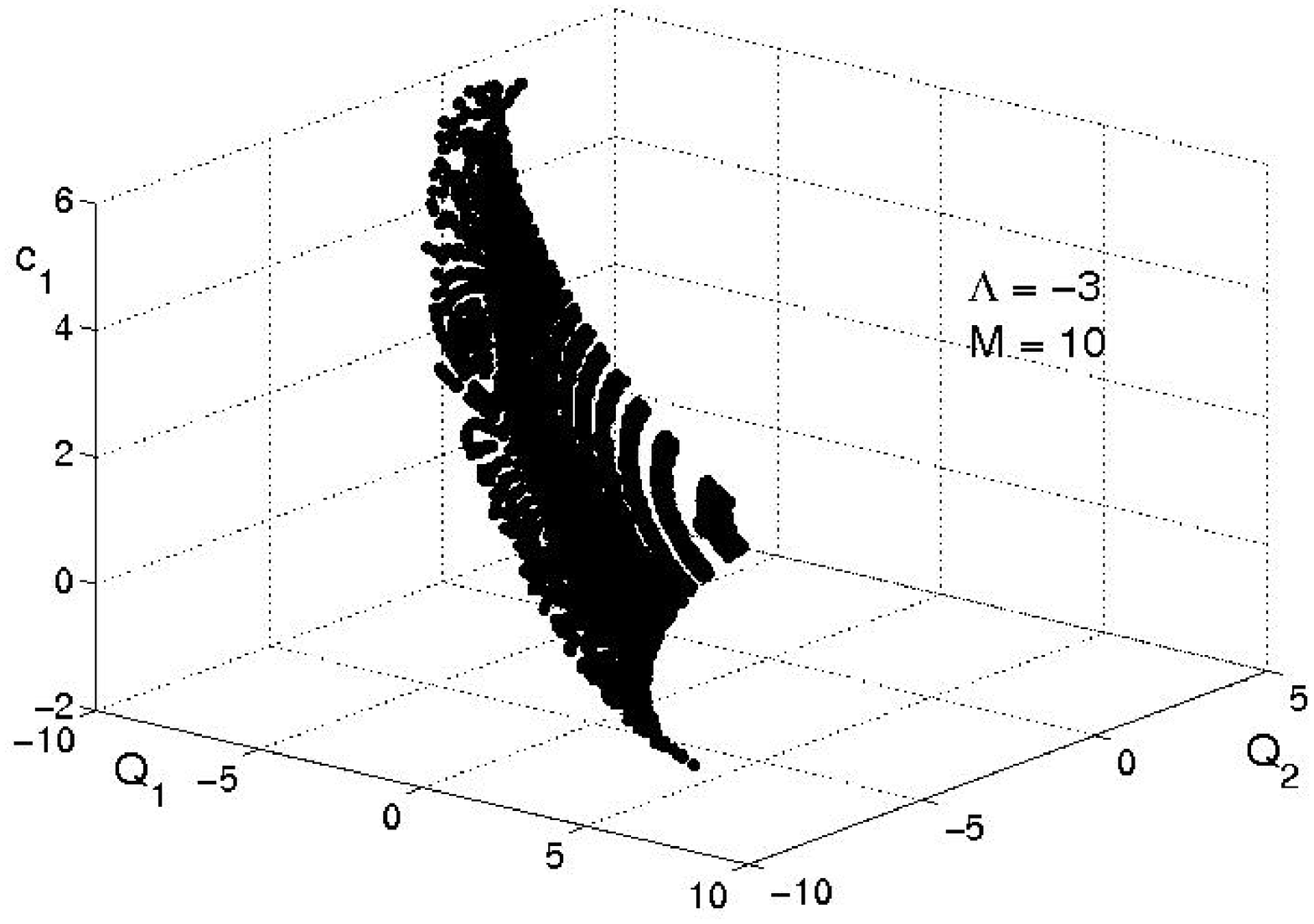}
\, \, \,
\includegraphics[angle=270,width=7cm]{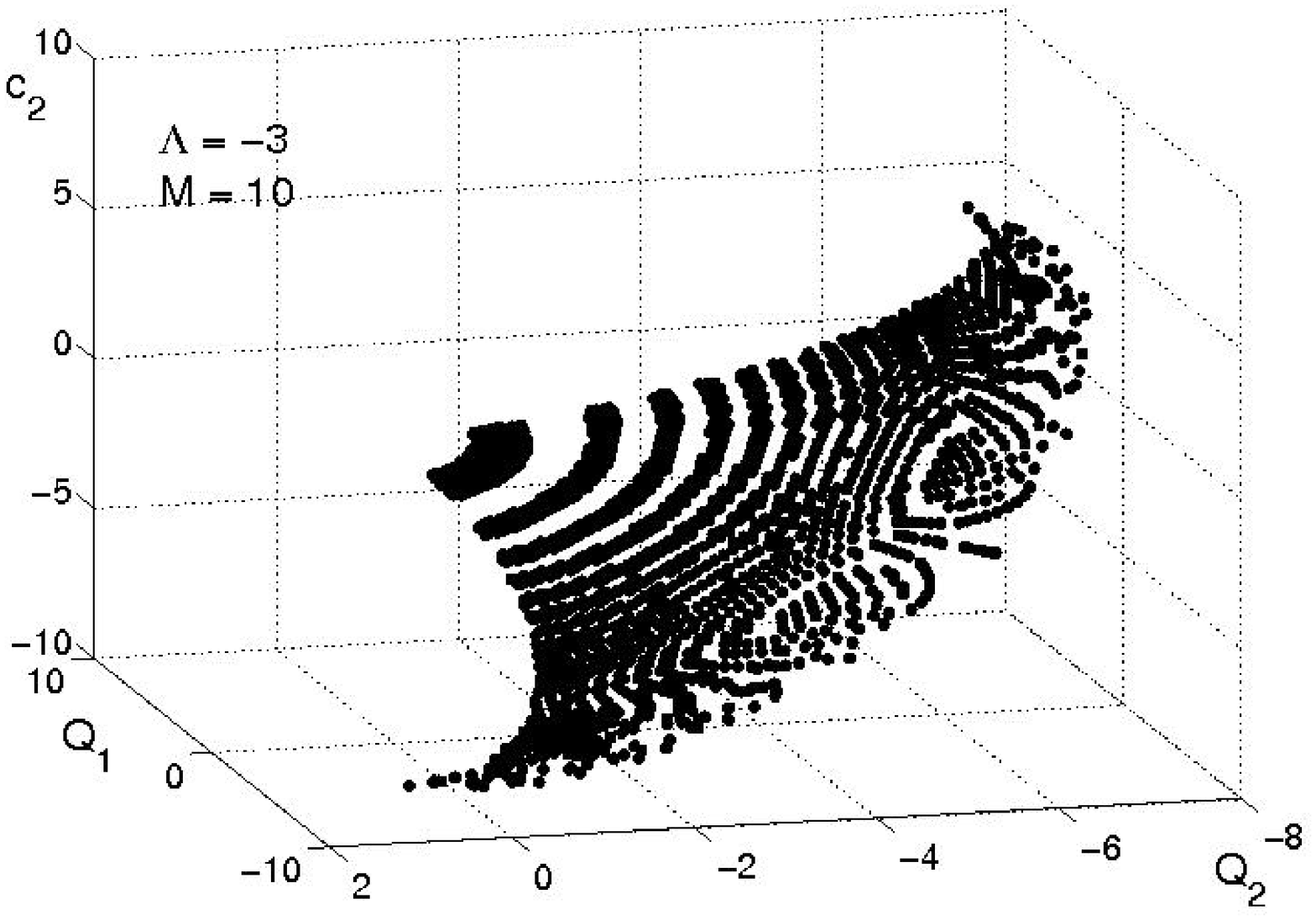}
\caption{Scatter plot of the constants $c_{1}$ (left) and $c_{2}$ (right)
against the non-Abelian charges $Q_{1}$ and $Q_{2}$ for the same black holes as in figure \ref{fig:su3q1q2}. It appears to be the case that $c_{1}$ and $c_{2}$ are single-valued functions of $Q_{1}$ and $Q_{2}$.}
\label{fig:su3q1q2c}
\end{center}
\end{figure}

For ${\mathfrak {su}}(3)$ and larger gauge groups, it is more difficult to see graphically whether the $c_{j}$ are
uniquely determined by $M$ and the charges $Q_{j}$, simply because there are more variables to plot.
However, all the numerical evidence we have been able to gather in the ${\mathfrak {su}}(3)$ case
does indicate that the $c_{j}$ are uniquely determined by
$M$ and $Q_{j}$.
To illustrate this, we consider ${\mathfrak {su}}(3)$ black holes with
$\Lambda = -3$, such that the inequalities (\ref{eq:stabineqs}) are satisfied\footnote{For ease of reference, we call these ``potentially stable'' black holes. We are not claiming to have shown that all black holes satisfying the inequalities (\ref{eq:stabineqs}) are stable. While the inequalities (\ref{eq:stabineqs}) are sufficient to guarantee stability in the sphaleronic perturbation sector,  they do not guarantee stability in the gravitational sector \cite{Baxter4}.}.
We would like to show that the $c_{1}$ and $c_{2}$ are determined by $M$, $Q_{1}$ and $Q_{2}$, where the non-Abelian charges
$Q_{1}$ and $Q_{2}$ are given by (\ref{eq:Qjfinal}):
\begin{equation}
Q_{1} = 1- \omega _{1, \infty}^{2} + \frac {1}{2} \omega _{2,\infty }^{2}, \qquad
Q_{2} = {\sqrt {3}} \left( 1 - \frac {1}{2} \omega _{2,\infty }^{2} \right) .
\label{eq:su3q1q2}
\end{equation}
To produce sensible plots, we first fix $M=10\pm0.1$ (there has to be some range of values of $M$ to obtain clear plots).
We perform a scan over the black hole solutions, numerically integrating the field equations for a grid of values of $r_{h}$,
$\omega _{1}(r_{h})$, and $\omega _{2}(r_{h})$.
The results are shown in figures \ref{fig:su3q1q2}--\ref{fig:su3q1q2c}.

The values of $Q_{1}$ and $Q_{2}$ for these black holes are shown in figure \ref{fig:su3q1q2}.
We observe that the values of the non-Abelian charges $Q_{1}$ and $Q_{2}$ have a band-like structure, with the bands curving round the
origin.   As the charges become more negative, the structure becomes less clear due to the finite grid on which we find our solutions.
We found similar behaviour for other values of $M$.

Scatter plots of $c_{1}$ and $c_{2}$ as functions of $Q_{1}$ and $Q_{2}$ for $M=10\pm 0.1$ can be found in figure \ref{fig:su3q1q2c}.
It is difficult to see clearly from figure \ref{fig:su3q1q2c}, but rotating the scatter plots on a computer screen
indicates that both $c_{1}$ and $c_{2}$ are single-valued functions of $Q_{1}$ and $Q_{2}$.
Again, we find similar behaviour for other values of $M$.
This provides some (albeit limited) numerical evidence for the parameters $c_{1}$ and $c_{2}$ being determined by the non-Abelian charges $Q_{1}$
and $Q_{2}$.

\begin{figure}
\begin{center}
\includegraphics[angle=270,width=8cm]{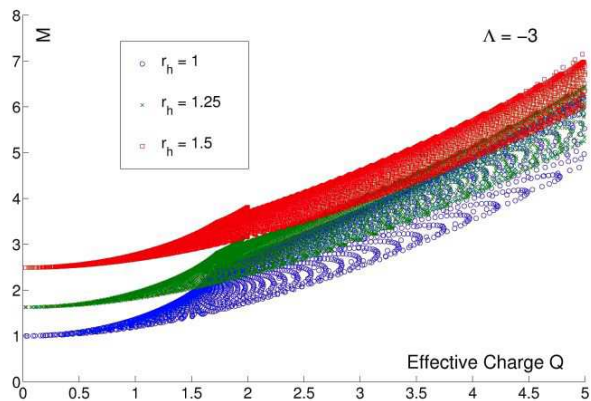}
\caption{Scatter plot of mass $M$ as a function of effective charge $Q$ (\ref{eq:effectivecharge}) for ${\mathfrak {su}}(3)$ black holes with
$\Lambda = -3$ and event horizon radius $r_{h}=1$ (black), $1.25$ (blue), $1.5$ (red).
It can be seen that the surfaces for different $r_{h}$ overlap, indicating that $M$ and $Q$ do not uniquely characterize the black hole solutions.}
\label{fig:su3mq}
\end{center}
\end{figure}

As in the ${\mathfrak {su}}(2)$ case, further numerical evidence that the mass and non-Abelian charges characterize the black holes can be found
by plotting mass $M$ as a function of the charges.
Firstly, in figure \ref{fig:su3mq}, we plot $M$ as a function of the effective charge $Q$ (\ref{eq:effectivecharge}), to see if the quantites $M$
and $Q$ are sufficient to uniquely fix the black hole solution.
In figure \ref{fig:su3mq}, we see that the surfaces corresponding to different values of the event horizon radius $r_{h}$ overlap.
For example, there are black holes with mass $M=6.1$ and effective charge $Q=5$ for $r_{h}=1$, $1.25$ and $1.5$.
We therefore conclude that ${\mathfrak {su}}(3)$ black holes are not uniquely specified by their mass $M$ and effective charge $Q$.

\begin{figure}[h]
\begin{center}
\includegraphics[angle=270,width=10cm]{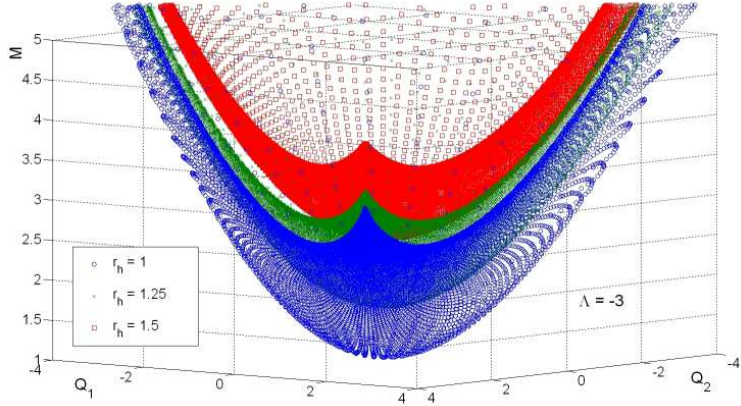}
\caption{Scatter plot of mass $M$ as a function of non-Abelian charges $Q_{1}$ and $Q_{2}$ (\ref{eq:su3q1q2}) for the same black hole solutions
as in figure \ref{fig:su3mq}. The different coloured surfaces are for different values of the event horizon radius $r_{h}$. There is no evidence that the surfaces for different $r_{h}$ intersect, suggesting that $M$, $Q_{1}$ and $Q_{2}$
uniquely characterize the black holes.}
\label{fig:su3mq1q2}
\end{center}
\end{figure}

On the other hand, if we plot $M$ as a function of the non-Abelian charges $Q_{1}$ and $Q_{2}$ (\ref{eq:su3q1q2}),
a different structure can be seen in figure \ref{fig:su3mq1q2}.
The surfaces in figure \ref{fig:su3mq1q2} appear to foliate the parameter space, and we found no evidence of the surfaces for
different $r_{h}$ intersecting.
This indicates that the parameters $M$, $Q_{1}$ and $Q_{2}$ do indeed uniquely characterize the black hole solutions.

\section{Analyticity of $M$, $Q_{j}$ and $c_{j}$ as functions of horizon parameters}
\label{sec:analyticity}

The numerical analysis of the previous section has indicated that the $N+1$
parameters $M$, $\Lambda $ and $Q_{j}$, for $j=1,\ldots N-1$, completely determine
stable black hole solutions.
In section \ref{sec:analytic}, we will give an analytic argument to show that this assertion holds, at least
for a large subspace of stable black holes.
The essence of the argument is to consider the $N+1$ parameters
$\left( \Lambda, M, Q_{1},\ldots ,Q_{N-1} \right) $ as functions of the
the $N+1$ parameters $\left( \Lambda ,r_{h},\omega _{1}(r_{h}), \ldots
,\omega _{N-1}(r_{h}) \right) $
which are known, from \cite{Baxter3}, to uniquely characterize
black hole solutions of the field equations.
If the map
\begin{equation}
\left( \Lambda ,r_{h},\omega _{1}(r_{h}), \ldots
,\omega _{N-1}(r_{h}) \right) \rightarrow
\left( \Lambda , M, Q_{1},\ldots ,Q_{N-1} \right)
\label{eq:Jmap}
\end{equation}
is invertible, then we can deduce that
$\left( \Lambda, M, Q_{1},\ldots, Q_{N-1} \right) $ uniquely characterize the black
hole solutions.
A convenient way to show that the map (\ref{eq:Jmap}) is invertible is to use the
inverse function theorem, so that if we can show that
the Jacobian $J$, given by
\begin{equation}
J = \left|
\begin{array}{ccccc}
1 & 0 & 0 & \cdots  & 0 \\
0 & {\displaystyle {\frac {\partial M}{\partial r_{h}} }}
& {\displaystyle {\frac {\partial M}{\partial \omega _{1}(r_{h})}  }} & \cdots &
{\displaystyle {\frac {\partial M}{\partial \omega _{N-1}(r_{h})} }}
\\
0 & {\displaystyle {\frac {\partial Q_{1}}{\partial r_{h}} }}
& {\displaystyle {\frac {\partial Q_{1}}{\partial \omega _{1}(r_{h})} }} & \cdots &
{\displaystyle {\frac {\partial Q_{1}}{\partial \omega _{N-1}(r_{h})} }}
\\
\vdots & \vdots & \vdots & \ddots & \vdots \\
0 &{\displaystyle { \frac {\partial Q_{N-1}}{\partial r_{h}} }} &
{\displaystyle {\frac {\partial Q_{N-1}}{\partial \omega _{1}(r_{h})} }} & \cdots &
{\displaystyle {\frac {\partial Q_{N-1}}{\partial \omega _{N-1}(r_{h})} }}
\end{array}
\right| ,
\label{eq:Jacobian}
\end{equation}
is non-vanishing, then the map (\ref{eq:Jmap}) is invertible.
However, this assumes that the map (\ref{eq:Jmap}) is continuously differentiable.
In fact, in this section we will be able to prove a stronger condition, namely
that the quantities $M$, $Q_{j}$ and $c_{j}$ (\ref{eq:infinity}) are
analytic functions of $\left( \Lambda ,r_{h},\omega _{1}(r_{h}), \ldots
,\omega _{N-1}(r_{h}) \right) $.

We begin by writing the field equations (\ref{eq:YMe}, \ref{eq:Ee}) in
an alternative form, following Proposition 7 in \cite{Baxter3}.
New variables $\lambda $, $\psi _{k}$ and $\xi _{k}$ are defined by
\begin{equation}
\lambda = \frac {r}{r_{h}} \left( 1 - \mu - \frac {\Lambda r^{2}}{3} \right),
\qquad
\psi _{j} = \omega _{j},
\qquad
\xi _{j}= \frac {r^{2}}{r_{h}} \frac {d\omega _{j}}{dr} ,
\label{eq:infnewvars}
\end{equation}
and then the field equations take the following form:
\begin{eqnarray}
& &
z\frac {d\lambda }{dz} = z f_{\lambda } , \quad
z\frac {d\psi _{j} }{dz} = -z\xi _{j}, \quad
z\frac {d\xi _{j}}{dz} = zf_{\xi _{j}}, \quad
\nonumber \\ & &
z\frac {dS}{dz} = z^{4} f_{S}, \quad
z\frac {d\Lambda }{dz} =0 , \quad
z\frac {dr_{h}}{dz} = 0,
\label{eq:infeqns}
\end{eqnarray}
where the new independent variable is $z = r_{h}/r$, so that $z\in [0,1]$.
The functions $f_{\lambda }$, $f_{\xi _{j}}$ and $f_{S}$ can be found in
\cite{Baxter3} (noting that in that reference, $\xi _{j}$ is $r_{h}$
times $\xi _{j}$ as defined in (\ref{eq:infnewvars}));
their detailed form is not essential here, their key property being
that they are analytic functions of $z$,
$\Lambda $, $r_{h}$, $\lambda $, the $\psi _{j}$
and the $\xi _{j}$, at least in a neighbourhood of $z=0$.
Proposition 7 of \cite{Baxter3} then gives local existence of solutions of the
differential equations (\ref{eq:infeqns}), in a neighbourhood of $z=0$,
these solutions being analytic in $M$, $\Lambda $, $r_{h}$, $\omega _{j,\infty }$
and ${\tilde {c}}_{j}=c_{j}/r_{h}$.

The equations (\ref{eq:infeqns}) are regular only on the interval $z\in [0,1) $.
In the proof of Proposition 6 in \cite{Baxter3}, equations of a similar form to
(\ref{eq:infeqns}), but with different independent and dependent variables, are
derived and used to show the existence of local solutions in a neighbourhood of the
event horizon, and analytic in $\Lambda $, $r_{h}$, and $\omega _{j}(r_{h})$.
These latter equations are regular on an interval equivalent to $z\in (0,1]$.
To show that $M$, $\omega _{j,\infty }$ and $c_{j}$ are analytic functions of
$r_{h}$, $\Lambda $ and $\omega _{j}(r_{h})$ we therefore have to match the solutions
which exist in a neighbourhood of the event horizon with those that exist in a
neighbourhood of infinity.

To do this, let ${\tilde {z}} = z-\frac {1}{2}$.  Then the equations
(\ref{eq:infeqns}) can be written in terms of this new independent variable as
\begin{eqnarray}
& &
{\tilde {z}}\frac {d\lambda }{d{\tilde {z}}} = {\tilde {z}} f_{\lambda } , \quad
{\tilde {z}}\frac {d\psi _{j} }{d{\tilde {z}}} = -{\tilde {z}}\xi _{j}, \quad
{\tilde {z}}\frac {d\xi _{j}}{d{\tilde {z}}} = {\tilde {z}}f_{\xi _{j}}, \quad
\nonumber \\ & &
{\tilde {z}}\frac {dS}{d{\tilde {z}}} = {\tilde {z}} z^{3} f_{S}, \quad
{\tilde {z}}\frac {d\Lambda }{d{\tilde {z}}} =0 , \quad
{\tilde {z}}\frac {dr_{h}}{d{\tilde {z}}} = 0.
\label{eq:ztilde}
\end{eqnarray}
Therefore, by analogy with (\ref{eq:infeqns}), we have a local existence theorem
for solutions of the above differential equations, in a neighbourhood of
${\tilde {z}}=0$ ($z=\frac {1}{2}$),
and the solutions are analytic in $r_{h}$, $\Lambda $, and the
values of $\lambda $, $S$, $\psi _{j}$ and $\xi _{j}$ at ${\tilde {z}}=0$.

Since the equations (\ref{eq:ztilde}) are obtained from (\ref{eq:infeqns})
by a simple translation of the independent variable, the new equations
(\ref{eq:ztilde}) are regular at ${\tilde {z}}=-\frac {1}{2}$, which corresponds
to $z=0$.
Therefore the values of the field variables at $z=0$ are analytic in $r_{h}$,
$\Lambda $ and the values of $\lambda $, $S$, $\psi _{j}$ and $\xi _{j}$
at ${\tilde {z}}=0$.
In particular, the quantities $M$, $Q_{j}$ and $c_{j}$ will be analytic in $r_{h}$,
$\Lambda $ and the values of $\lambda $, $S$, $\psi _{j}$ and $\xi _{j}$
at ${\tilde {z}}=0$.

Now suppose we have a space of black hole solutions of the original field equations
(\ref{eq:YMe}, \ref{eq:Ee}) which are regular everywhere between the event horizon
and infinity.
In particular, every member of this space of solutions is regular at
$z=0$ (infinity),  $z=\frac {1}{2}$ and $z=1$ (event horizon).
From Proposition 6 in \cite{Baxter3}, these solutions are analytic in $r_{h}$,
$\Lambda $, and $\omega _{j}(r_{h})$ in the neighbourhood of the event horizon
in which they exist. Therefore, the values of the field variables at $z=\frac {1}{2}$
(${\tilde {z}}=0$) are analytic in $r_{h}$, $\Lambda $ and $\omega _{j}(r_{h})$.

We now have the following situation: the values of the $\lambda $, $S$,
$\psi _{j}$ and $\xi _{j}$ at ${\tilde {z}}=0$
are analytic functions of $r_{h}$, $\Lambda $
and $\omega _{j}(r_{h})$.
In addition, the quantities $M$, $Q_{j}$ and $c_{j}$ are analytic in
$r_{h}$, $\Lambda $ and the values of $\lambda $, $S$, $\psi _{j}$ and
$\xi _{j}$ at ${\tilde {z}}=0$.
Therefore we can conclude that $M$, $Q_{j}$ and $c_{j}$ are analytic in $r_{h}$,
$\Lambda $ and the $\omega _{j}(r_{h})$.
Furthermore, as a corollary, it must be the case that the Jacobian $J$ (\ref{eq:Jacobian}) is also an analytic function of $r_{h}$, $\Lambda $ and the
$\omega _{j}(r_{h})$.

\section{Characterizing ${\mathfrak {su}}(N)$ EYM black holes for large
$\left| \Lambda \right| $ - analytic work}
\label{sec:analytic}

Our purpose in this section is to argue that the charges $Q_{j}$ (\ref{eq:Qjfinal}),
together with
the mass $M$ and cosmological constant $\Lambda $, are sufficient to
uniquely characterize ${\mathfrak {su}}(N)$ hairy black holes, at least for
some subset of those black holes which are linearly stable \cite{Baxter4}.

There are a number of difficulties in deriving such a result: firstly,
general black hole solutions are known only numerically;
secondly, while the local existence theorems in \cite{Baxter3} near
the event horizon
and at infinity are valid for any values of the cosmological constant,
event horizon
radius and the other parameters in the theory,
the theorems proving the existence of regular black hole solutions are valid
only for sufficiently large $\left| \Lambda \right| $.

In other words, in \cite{Baxter3} the existence of black hole solutions
is shown for
fixed $r_{h}$ and $\omega _{j}(r_{h})$, and sufficiently large
$\left| \Lambda \right| $,
where, as discussed in \cite{Baxter3}, how large ``sufficiently large''
is will likely
depend on the values of the parameters $r_{h}$ and $\omega _{j}(r_{h})$.
In the present analysis we wish to study the problem from a different
perspective:
we want to fix $\left| \Lambda \right| $ to be some (suitably large) value,
and
vary $r_{h}$ and $\omega _{j}(r_{h})$, and then show that, at least
for some subset
of the black hole solutions thus generated, the black holes can be
uniquely
characterized by their mass $M$ and non-Abelian charges $Q_{j}$.

From our above discussion of the generalized ``no-hair'' conjecture,
it is clear that
we are only interested in characterizing stable hairy black hole solutions.
Due to the symmetry (\ref{eq:omegaswap}) of the field equations,
we may consider only $\omega _{j} (r_{h})>0$ without loss of generality.
We therefore restrict attention to black holes for which the inequalities
(\ref{eq:stabineqs}) are satisfied, both on and outside the event horizon.
It is straightforward to show in this case that the functions $\omega _{j}(r)$
are monotonically increasing and that $\omega _{j}(r)>0$ for all $r\ge r_{h}$.
We also consider only those black holes which are thermodynamically
stable (see section \ref{sec:thermodynamics}), which is equivalent to focussing
attention on comparatively
large black holes.

\subsection{Characterizing black holes by $M$ and $Q_{j}$}
\label{sec:bhs}

We begin our analysis with a more careful consideration of the space of black hole
solutions for large $\left| \Lambda \right| $.
There are two length scales in this problem: the event horizon radius $r_{h}$
and $\ell = {\sqrt {-3/\Lambda }}$.
We wish to study black hole solutions when $\ell $ is small,
but allowing $r_{h}$ to
vary.
To this end, it is helpful to introduce a dimensionless
radial co-ordinate $x=r/r_{h}$,
so that $x \in [1, \infty ) $ irrespective of the value of the
event horizon radius
$r_{h}$.
In terms of $x$ the field equations (\ref{eq:YMe}, \ref{eq:Ee}) take the form:
\begin{eqnarray}
\frac {d{\hat {m}}}{dx} & = &
\mu G + r_{h}^{2} x^{2} p_{\theta } ;
\nonumber
\\
\frac {1}{S} \frac {dS}{dx} & = & \frac {2G}{x} ;
\nonumber
\\
0 & = &
x^{2} \mu \frac {d^{2}\omega _{j}}{dx^{2}} +
\left[ 2{\hat {m}} - 2r_{h}^{2} x^{3} p_{\theta }
+ \frac {2r_{h}^{2}x^{3}}{\ell ^{2}}
\right] \frac {d\omega _{j}}{dx} + W_{j} \omega _{j} ;
\label{eq:nondim}
\end{eqnarray}
where ${\hat {m}}(x) = m(r)/r_{h}$, the quantity $W_{j}$ is
given by (\ref{eq:Wdef})
and equations (\ref{eq:ptheta}, \ref{eq:Gdef}) now take the form
\begin{eqnarray}
p_{\theta } & = &
\frac {1}{4x^{4}r_{h}^{4}} \sum _{j=1}^{N}
\left[\left(\omega^2_j-\omega^2_{j-1}-N-1+2j\right)^2\right] ;
\nonumber \\
G & = &
\frac {1}{r_{h}^{2}} \sum _{j=1}^{N-1}
\left( \frac {d\omega _{j}}{dx} \right) ^{2}.
\end{eqnarray}
At the event horizon, $x=1$, we have
\begin{equation}
{\hat {m}}(1) = \frac {1}{2}
\left( 1 + \frac {r_{h}^{2}}{\ell ^{2}} \right) ,
\label{eq:hatm1}
\end{equation}
which becomes large as $\ell \rightarrow 0$.
We define a further new variable ${\tilde {m}}(x)$ by
\begin{equation}
{\hat {m}}(x) = m_{1} + {\tilde {m}}(x)
\end{equation}
where $m_{1} = {\hat {m}}(1)$ (\ref{eq:hatm1}).
In \cite{Baxter3}, it is shown that
$\frac {d}{dx} \left( {\tilde {m}}\ell ^{2} \right)
\rightarrow 0 $ as $\ell \rightarrow 0$
for fixed $r_{h}$, $\omega _{j}(r_{h})$,
but this does not necessarily mean that
$\frac {d{\tilde {m}}}{dx}$ tends to zero in
this limit.
Indeed, we shall see below that this is not the case,
as is borne out by Figures
5 and 6 in \cite{Baxter3}.

We now write the first and third field equations (\ref{eq:nondim}) in the form
\begin{eqnarray}
\ell ^{2} \frac {d{\tilde {m}}}{dx}
& = &
\frac {1}{r_{h}^{2}} \left[ \ell ^{2}
- \frac {2m_{1}\ell ^{2}}{x} - \frac {2{\tilde {m}}\ell ^{2}}{x}
+ r_{h}^{2} x^{2} \right]
\sum _{j=1}^{N-1} \left( \frac {d\omega _{j}}{dx} \right) ^{2}
\nonumber \\ & &
+ \frac {\ell ^{2}}{4x^{2}r_{h}^{2}}
\sum _{j=1}^{N}
\left[\left(\omega^2_j-\omega^2_{j-1}-N-1+2j\right)^2\right] ;
\nonumber
\\
0 & = &
x^{2} \left[ \ell ^{2} - \frac {2m_{1}\ell ^{2}}{x} -
\frac {2{\tilde {m}}\ell ^{2}}{x} + r_{h}^{2}x^{2} \right]
\frac {d^{2}\omega _{j}}{dx^{2}}
\nonumber \\ & &
+ \left[ 2m_{1}\ell ^{2} + 2{\tilde {m}}\ell ^{2} - 2r_{h}^{2}x^{3} \ell ^{2}
p_{\theta }
+ 2r_{h}^{2} x^{3} \right] \frac {d\omega _{j}}{dx}
+ \ell ^{2} W_{j} \omega _{j} .
\label{eq:smallell}
\end{eqnarray}
The analysis leading to Proposition 11 in \cite{Baxter3}
essentially involves ignoring the terms
$ {\tilde {m}} \ell ^{2}$ and $\ell ^{2} W_{j}\omega _{j}$ in the right-hand-sides of
the above equations.
It is clear that for any {\em {fixed}} values of $r_{h}$ and $\omega _{j}(r_{h})$,
we may
choose $\ell $ sufficiently small that these terms are negligible compared with the
others in the equations.
However, here we wish to vary $r_{h}$ and $\omega _{j}(r_{h})$, and therefore
we need to carefully examine the magnitudes of all the quantities in the above
equations for small $\ell $.
It is clear from equations (\ref{eq:smallell})
that there are subtleties if $r_{h}$
is also small.
For this reason,
we  consider only those black holes
for which $r_{h}\gg \ell $, as these will be thermodynamically stable
(see section \ref{sec:thermodynamics}).

For a regular event horizon at $x=1$, we can vary $r_{h}$ and $\omega _{j}(r_{h})$
subject to the single constraint (\ref{eq:constraint}), which can be written as
\begin{equation}
\ell ^{2} \sum _{j=1}^{N}
\left[\left(\omega^2_j(r_{h})-\omega^2_{j-1}(r_{h})-N-1+2j\right)^2\right]
< 2r_{h}^{2}\ell ^{2} + 6r_{h}^{4} .
\label{eq:wconstraint}
\end{equation}
Therefore, for each $j$, we have
\begin{equation}
\ell ^{2} \left(\omega^2_j(r_{h})-\omega^2_{j-1}(r_{h})-N-1+2j\right)^2
< 2r_{h}^{2}\ell ^{2} + 6r_{h}^{4}
\end{equation}
and, as a result, it must be the case that
\begin{equation}
\ell \left[ \omega _{j}^{2}(r_{h}) -j \left( N-j \right) \right]
 < j \left[ 2r_{h}^{2}\ell ^{2} + 6r_{h}^{4} \right] ^{\frac {1}{2}} .
\end{equation}
Our numerical results in section \ref{sec:numerics} indicate that we do not have
regular black hole solutions for values of $\omega _{j}(r_{h})$ close to the
boundary of the region defined by (\ref{eq:wconstraint}), and therefore we do
not need to consider all $\omega _{j}(r_{h})$ such that (\ref{eq:wconstraint})
is satisfied.
At the same time, we find that the region of parameter space for which we have
regular black hole solutions which are uniquely determined by $M$ and $Q_{j}$
seems to grow as $\ell $ decreases.
Therefore we would like to consider a region of the $\omega _{j}(r_{h})$
parameter space which, for small $\ell $, is smaller than the region
defined by the inequality (\ref{eq:wconstraint}), but which nonetheless grows
as $\ell $ decreases.

To this end, define new functions $q_{j}(x)$ by
\begin{equation}
\ell ^{2} \left[ \omega _{j}^{2}(r) -j \left( N-j \right) \right] ^{2}
= \ell ^{2\sigma } q_{j}^{2}(x) ,
\label{eq:qdef}
\end{equation}
where $\sigma >0$ is a presently unknown constant, and is the same for all $j$.
We further assume that $q_{j}(x)$ is order one for small $\ell $.
Setting $\sigma = 1$ would correspond to an upper bound on $\omega _{j}$
which is fixed independent of $\ell $, whilst setting $\sigma =0$ corresponds to
considering the whole of the region of
parameter space satisfying (\ref{eq:wconstraint}).
We therefore anticipate that $0<\sigma <1$ will be relevant for our analysis.
In this case, there is an upper bound on $\omega _{j}^{4}$ of the order of
$\ell ^{2\sigma - 2}q_{j}^{2}$ which grows as $\ell $ decreases.

In the ${\mathfrak {su}}(2)$ case,
it is shown in \cite{Winstanley1} that, for fixed $r_{h}$ and $\omega (r_{h})$,
$\ell ^{-1} \omega '(r) \rightarrow 0$ as $\ell \rightarrow 0$.
This suggests the following definition of further new functions $\eta _{j}(x)$,
again expected to be order one for small $\ell $, such that
\begin{equation}
\ell ^{-1} \frac {d\omega _{j}}{dx} = \ell ^{\kappa }\eta _{j}(x)
\label{eq:etadef}
\end{equation}
for some $\kappa >0$.

Let us now examine whether it is possible to find suitable values of $\sigma $
and $\kappa $ so as to obtain consistent approximate solutions of the field
equations (\ref{eq:smallell}) when $\ell $ is small.
First write the equations (\ref{eq:smallell}) in terms
of the $q_{j}$ and $\eta _{j}$:
\begin{eqnarray}
\ell ^{2} \frac {d{\tilde {m}}}{dx}
& = &
\frac {\ell ^{2\kappa + 2}}{r_{h}^{2}} \left[ \ell ^{2}
- \frac {2m_{1}\ell ^{2}}{x} - \frac {2{\tilde {m}}\ell ^{2}}{x}
+ r_{h}^{2} x^{2} \right]
\sum _{j=1}^{N-1} \eta _{j}^{2}(x)
\nonumber \\ & &
+ \frac {\ell ^{2\sigma }}{4x^{2}r_{h}^{2}}
\sum _{j=1}^{N}
\left[q_{j}(x) - q_{j-1}(x)\right] ^{2} ;
\label{eq:smallellm}
\\
0 & = &
x^{2}\ell ^{\kappa +1} \left[ \ell ^{2} - \frac {2m_{1}\ell ^{2}}{x} -
\frac {2{\tilde {m}}\ell ^{2}}{x} + r_{h}^{2}x^{2} \right]
\frac {d\eta _{j}}{dx}
\nonumber \\ & &
+\ell ^{\kappa + 1}
\left[ 2m_{1}\ell ^{2} + 2{\tilde {m}}\ell ^{2}
 - 2r_{h}^{2}x^{3} \ell ^{2} p_{\theta }
+ 2r_{h}^{2} x^{3} \right] \eta _{j}(x)
\nonumber \\ & &
+ \frac {1}{2}\ell ^{\sigma +1 } \left[ q_{j+1}(x)-2q_{j}(x)+q_{j-1}(x) \right]
\omega _{j}(x) .
\label{eq:smallellomega}
\end{eqnarray}
Examining first (\ref{eq:smallellomega}), and bearing in mind the upper bound
$\ell ^{2\sigma - 2}q_{j}^{2}$ on $\omega _{j}^{4}(x)$, for non-trivial solutions
we require that the first two terms are of the same order in $\ell $ as the last term.
This means that
\begin{equation}
\kappa  = \frac {3\sigma }{2} - \frac {1}{2}.
\label{eq:lambdafix}
\end{equation}
Requiring that $\kappa >0$ then implies that $\sigma > 1/3$, which is consistent
with our assumptions.
With this value of $\kappa $, we then have
\begin{equation}
2\kappa + 2 = 3\sigma +1 > 2\sigma
\end{equation}
for all $\sigma >0$, which means that the first line of (\ref{eq:smallellm})
is of subleading order in $\ell $ compared with the second.
Futhermore, differentiating (\ref{eq:qdef}) gives
\begin{equation}
\frac {dq_{j}}{dx} \sim 2q_{j} \eta _{j} \ell ^{1+ \sigma }
\end{equation}
for $\omega _{j}^{4} \sim \ell ^{2\sigma -2} q_{j}^{2}$ and $\kappa $ given by
(\ref{eq:lambdafix}).
Therefore we may regard the functions $q_{j}(x)$ as constant to leading order
in $\ell $.
Integrating (\ref{eq:smallellm}) then gives, to leading order in $\ell $,
\begin{eqnarray}
\ell ^{2} {\tilde {m}}(x) & = &  \frac {\ell ^{2\sigma }}{4r_{h}^{2}}
\left( 1 - \frac {1}{x} \right)
\sum _{j=1}^{N}
\left[q_{j}(1) - q_{j-1}(1)\right] ^{2}
\nonumber \\
& = & \frac {\ell ^{2}}{4r_{h}^{2}}
\left( 1 - \frac {1}{x} \right)
\sum _{j=1}^{N}
\left[\left(\omega^2_j(r_{h})-\omega^2_{j-1}(r_{h})-N-1+2j\right)^2\right],
\label{eq:mtildesmallell}
\end{eqnarray}
where we have used the initial condition ${\tilde {m}}=0$ at the event horizon $x=1$.
Note that the answer (\ref{eq:mtildesmallell}) implies that the terms
$\ell ^{2} {\tilde {m}}$ in (\ref{eq:smallellomega}) are indeed small compared with
$\ell ^{2} m_{1}= {\cal {O}}(1)$ for small $\ell $
and so can be ignored to first order.
However, it is not necessarily the case that ${\tilde {m}}$ itself is small.
We have ${\tilde {m}} \sim {\cal {O}} \left( \ell ^{2\sigma -2} \right) $,
so ${\tilde {m}}$ will in fact diverge as $\ell \rightarrow 0$,
albeit more slowly than
$m_{1}$ (\ref{eq:hatm1}).

To leading order, the Yang-Mills equations (\ref{eq:smallellomega}) become
\begin{eqnarray}
0 & = &
x^{2} \left[ r_{h}^{2}x^{2} - \frac {2m_{1}\ell ^{2}}{x}  \right]
\frac {d\eta _{j}}{dx}
+
\left[ 2m_{1}\ell ^{2}
+ 2r_{h}^{2} x^{3} \right] \eta _{j}(x)
\nonumber \\ & &
+ \frac {1}{2}\left[ q_{j+1}(x)-2q_{j}(x)+q_{j-1}(x) \right] q_{j}(x)^{\frac {1}{2}}
,
\label{eq:reducedYM}
\end{eqnarray}
where we have ignored the term
\begin{equation}
2r_{h}^{2}x^{3} \ell ^{2} p_{\theta } = 2r_{h}^{2} x^{3} \ell ^{2\sigma }
\sum _{j=1}^{N}
\left[q_{j}(x) - q_{j-1}(x)\right] ^{2},
\end{equation}
and used the leading order approximation
\begin{equation}
\omega _{j} = \ell ^{\frac {1}{2}\left( \sigma -1 \right) }q_{j}^{\frac {1}{2}}.
\end{equation}
To leading order in $\ell $, we can treat the functions
$q_{j}$ as approximately
constant, and in this case (\ref{eq:reducedYM}) can be integrated to give
\begin{equation}
\eta _{j} (x) =
-\frac {1}{2r_{h}^{2}\left( x^{2}+x+1 \right)}
\left[ q_{j+1}(1)-2q_{j}(1)+q_{j-1}(1) \right]
q_{j}(1)^{\frac {1}{2}},
\label{eq:etasolve}
\end{equation}
where we have chosen the arbitrary constant of
integration to be such that $\eta (x)$
is finite at the event horizon $x=1$.
Restoring the original variables, it is straightforward
to check that the solution
(\ref{eq:etasolve}) for $\eta _{j}$ satisfies
(\ref{eq:domegah}) at the event horizon.

Therefore we have obtained a consistent, approximate set of solutions of the
field equations which are valid for all $r_{h} \gg \ell $ and
all $\omega _{j}(r_{h})$
such that
\begin{equation}
\left[ \omega _{j}^{2}(r_{h}) -j \left( N-j \right) \right] ^{2}
< \ell ^{2\sigma -2}
\end{equation}
for some $\sigma \in \left( \frac {1}{3}, 1 \right) $.

\begin{figure}[h]
\begin{center}
\includegraphics[width=6cm,angle=270]{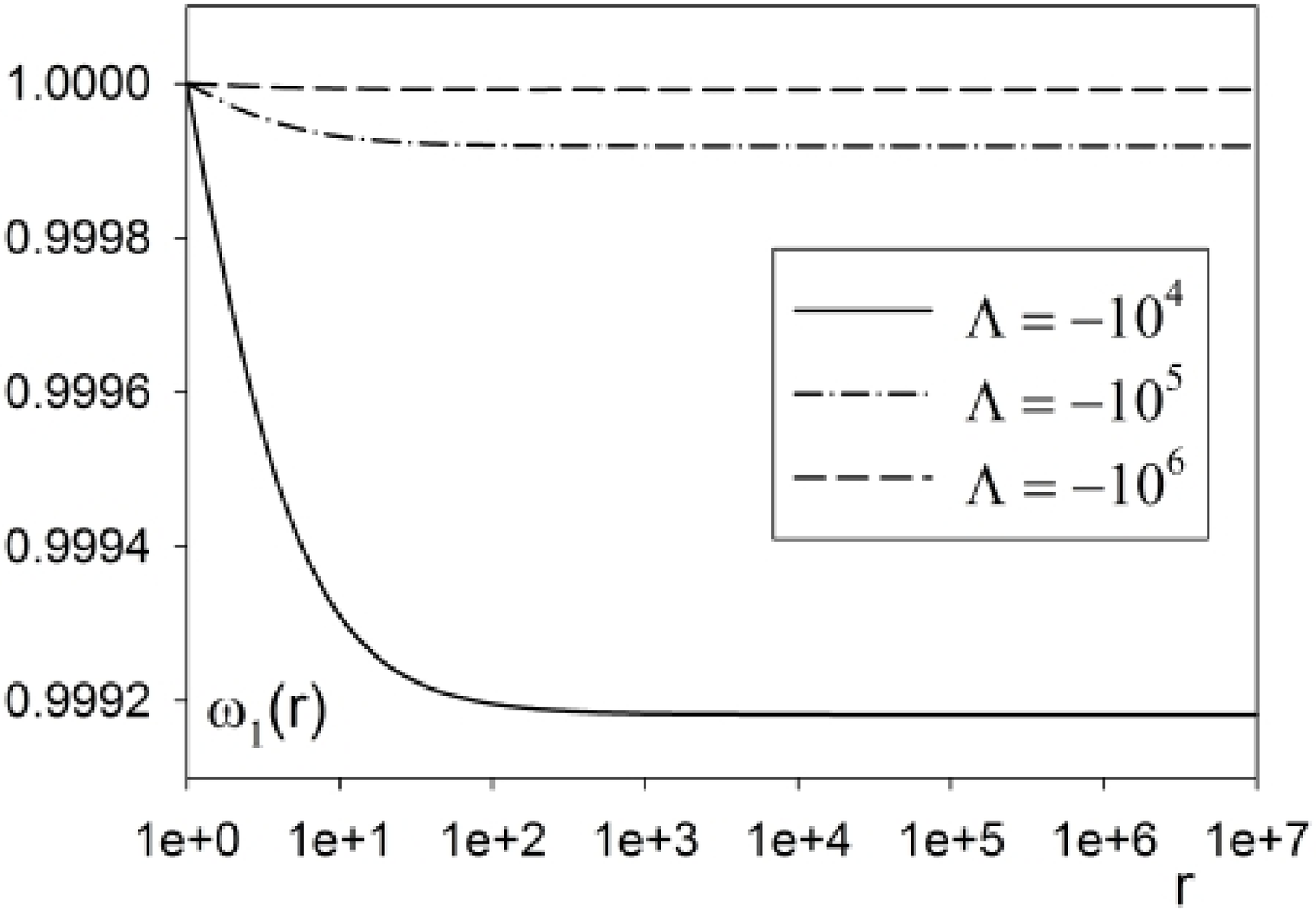}
\includegraphics[width=6cm,angle=270]{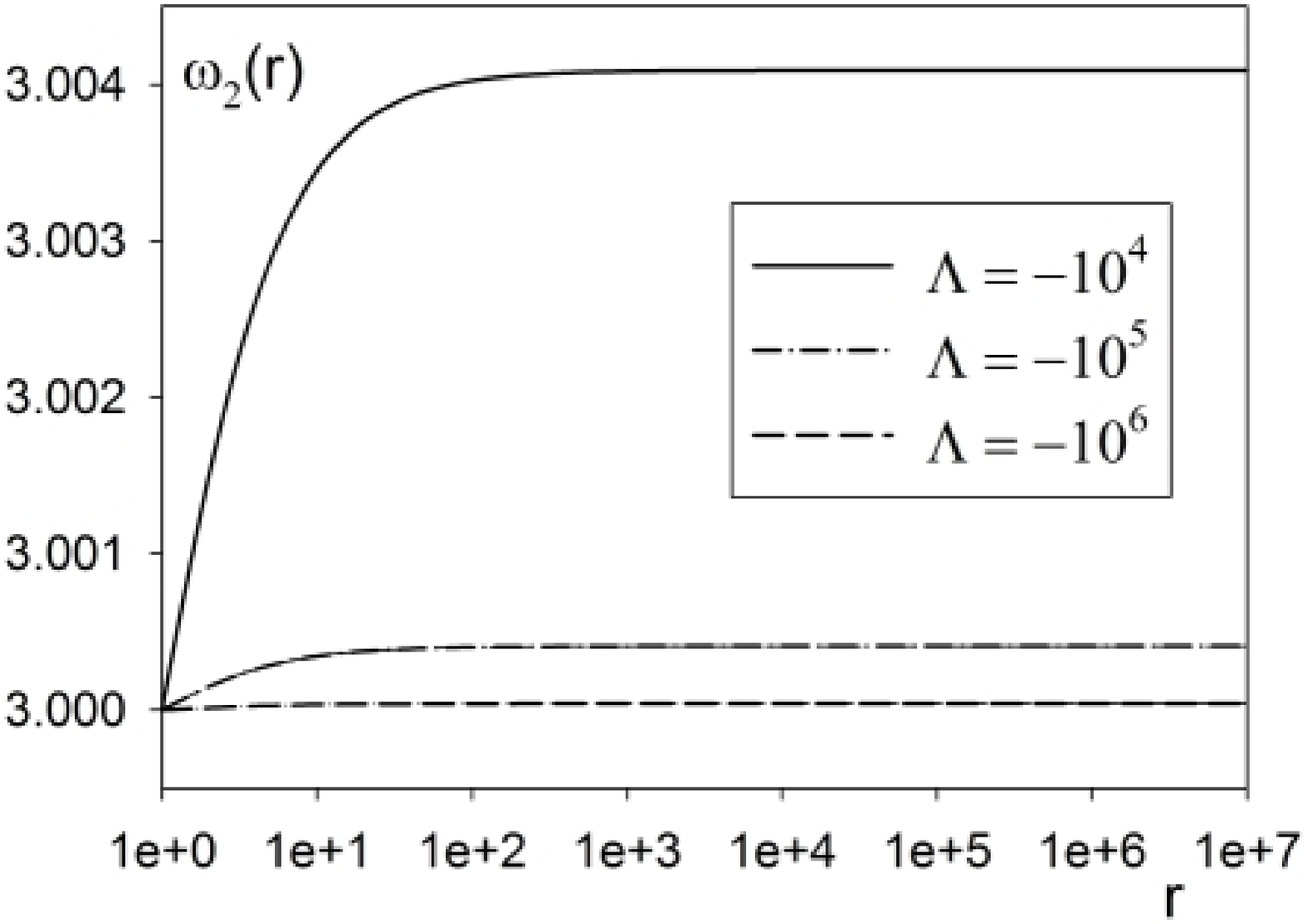}
\caption{Gauge field functions $\omega _{1}(r)$ (left) and $\omega _{2}(r)$ (right) for ${\mathfrak {su}}(3)$ black holes with
$r_{h}=1$, $\omega _{1}(r_{h})=1$, $\omega _{2}(r_{h})=3$ and varying cosmological constant $\Lambda =-10^{4}$, $-10^{5}$ and $-10^{6}$.
The gauge field functions approach constants as $\left| \Lambda \right|$ increases.}
\label{fig:largeLw}
\end{center}
\end{figure}

\begin{figure}[h]
\begin{center}
\includegraphics[width=5.5cm,angle=270]{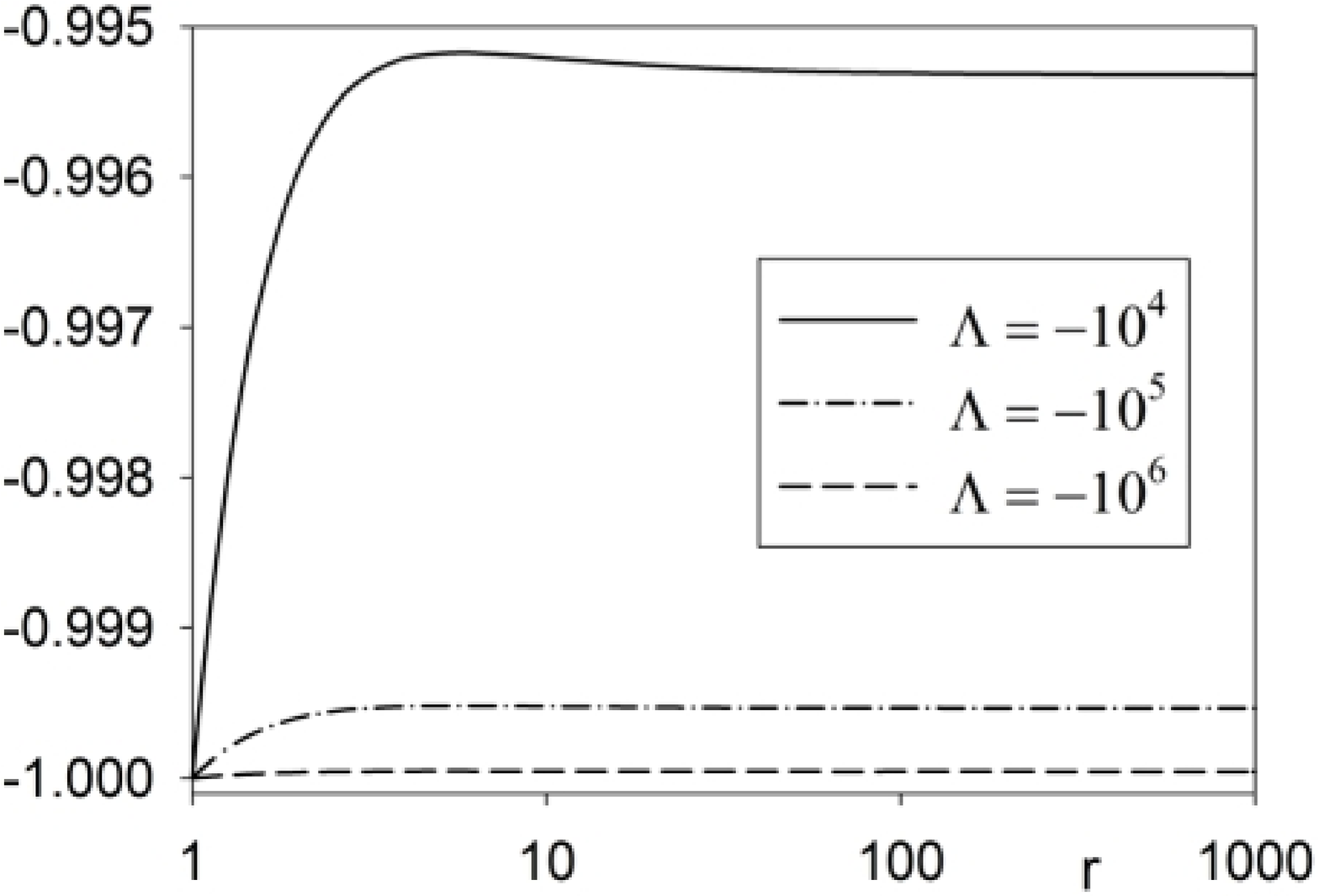}
\includegraphics[width=5.5cm,angle=270]{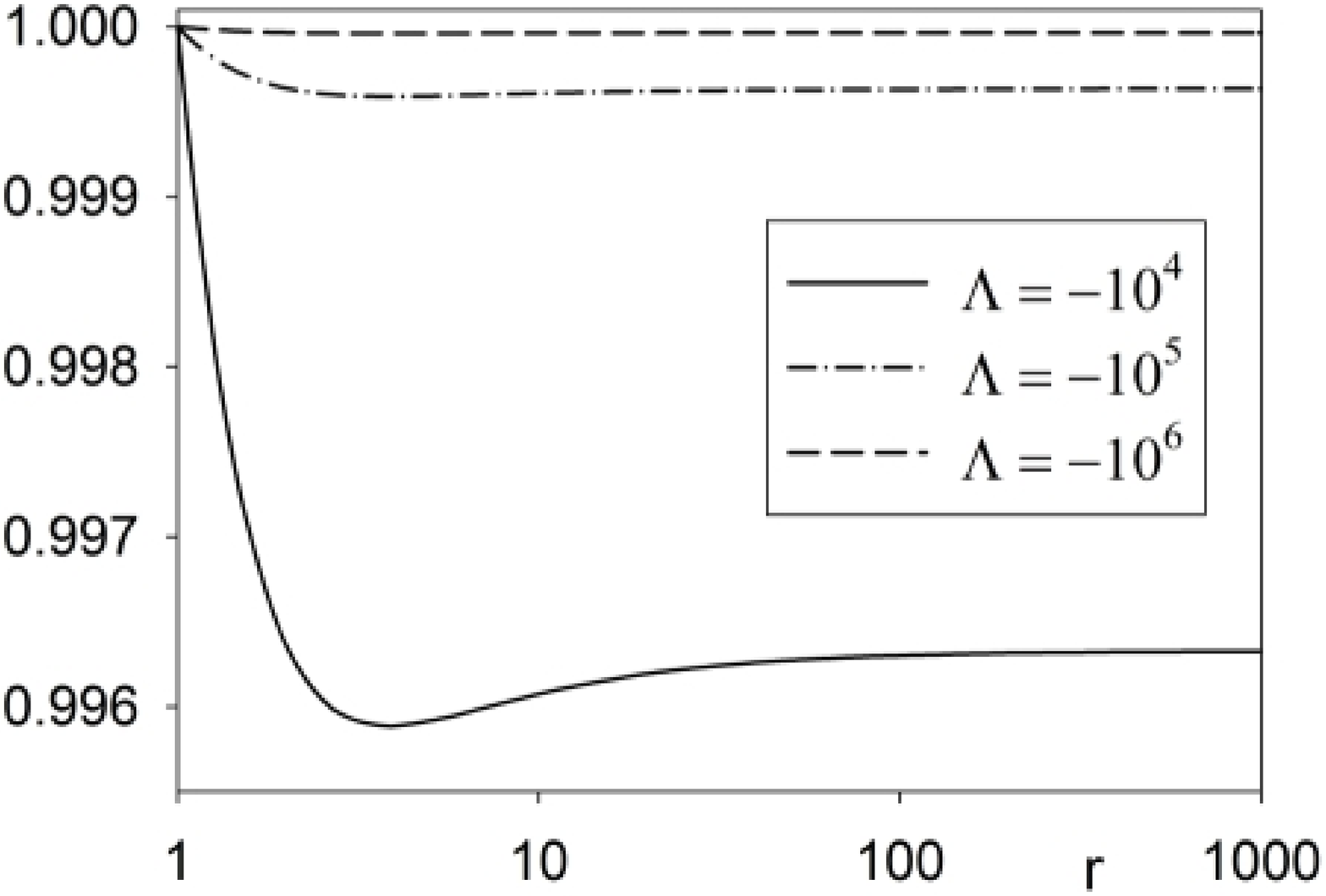}
\caption{Test of the approximation (\ref{eq:etasolve}) for $\omega _{1}'(r)$ (left) and $\omega _{2}'(r)$ (right). The numerical values of
$\omega _{j}'(r)$ for the black hole solutions in figure \ref{fig:largeLw} are divided by the approximate solution $\eta _{j}$ (\ref{eq:etasolve})
and a normalization constant, chosen so that the curves pass through $\pm 1$ at the event horizon.
The curves tend to $\pm 1$ for all $r$ as $\left| \Lambda \right|$ increases, indicating the validity of the approximation
$\eta _{j}$ (\ref{eq:etasolve}).}
\label{fig:largeLeta}
\end{center}
\end{figure}

\begin{figure}[h]
\begin{center}
\includegraphics[width=7cm,angle=270]{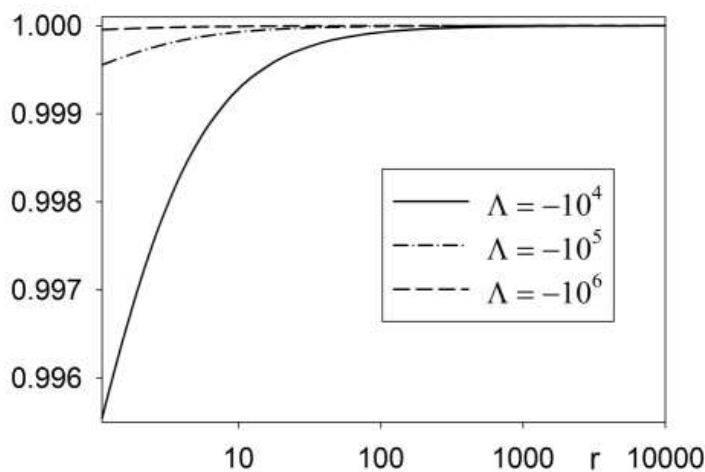}
\caption{Test of the approximation (\ref{eq:mtildesmallell}) for $m(r)-m(r_{h})$.  The numerical values of $m(r)-m(r_{h})$
for the black hole solutions in figure \ref{fig:largeLw} are divided by the approximate solution for ${\tilde {m}}$ (\ref{eq:mtildesmallell}),
and a normalization constant, chosen so that the curves pass through $1$ as $r\rightarrow \infty $.
The curves tend to $1$ for all $r$  as $\left| \Lambda \right|$ increases, indicating the validity of the approximation
 for $m(r)-m(r_{h})$ (\ref{eq:mtildesmallell}).}
\label{fig:largeLm}
\end{center}
\end{figure}

At this stage we compare our approximate solutions $q_{j} \approx $ constant,
$\eta _{j}$ (\ref{eq:etasolve}) and ${\tilde {m}}$ (\ref{eq:mtildesmallell}) with numerical solutions to test the validity of our approximations.
To illustrate the behaviour, in figures \ref{fig:largeLw}--\ref{fig:largeLm}, we consider ${\mathfrak {su}}(3)$ black holes with $r_{h}=1$,
$\omega _{1}(r_{h})=1$, $\omega _{2}(r_{h})=3$ and varying cosmological constant $\Lambda =-10^{4}$, $-10^{5}$ and $-10^{6}$.

Firstly, in figure \ref{fig:largeLw} we plot the gauge field functions $\omega _{1}(r)$ and $\omega _{2}(r)$.
It is clear that, as expected, the gauge field functions approach constants as $\left| \Lambda \right| \rightarrow \infty $.

To test the validity of the approximate solution for $\eta _{j}$ (\ref{eq:etasolve}), for each value of $r$ we divide the numerically-generated
value of $\omega _{j}'(r)$ by $\eta _{j}$ given by (\ref{eq:etasolve}).  The results are plotted in figure \ref{fig:largeLeta}, where we have also divided the answers by a constant (corresponding to the powers of $\ell $ in the definition (\ref{eq:etadef})), so that all curves pass through $\pm 1$ at the event horizon, to make comparisons easier.
Since $\omega _{1}'(r)<0$ and $\omega _{2}'(r)>0$ for the particular black hole solutions we are considering, we have fixed the additional constant
so that the curves for the
$\omega _{1}'(r)$ approximation to pass through $-1$ at the event horizon, and those for the $\omega _{2}'(r)$ approximation pass through $1$ at the
event horizon.
In figure \ref{fig:largeLeta}, we see that the curves tend to $\pm 1$ for all $r$ as $\left| \Lambda \right| \rightarrow \infty $.
This means that, in the large $\left| \Lambda \right|$ limit, the approximation $\eta _{j}$ (\ref{eq:etasolve}) becomes increasingly accurate.

To test the approximation ${\tilde {m}}$ (\ref{eq:mtildesmallell}), we take the numerical values of $m(r)-m(r_{h})$, divide by the approximation
(\ref{eq:mtildesmallell}) and, as in figure \ref{fig:largeLeta}, we also divide through by a constant (corresponding to powers of $\ell $) so that
all the curves pass through $1$ as $r\rightarrow \infty $.
We fix the normalization at $r\rightarrow \infty $ rather than at the event horizon because, as $r\rightarrow r_{h}$, both $m(r)-m(r_{h})$ and
${\tilde {m}}$ (\ref{eq:mtildesmallell}) vanish and numerical errors become an issue in dividing these two quantities.
The results are plotted in figure \ref{fig:largeLm}.
In figure \ref{fig:largeLm} we see that the curves tend to $1$ for all $r$ as $\left| \Lambda \right| \rightarrow \infty $,
showing that, in the large $\left| \Lambda \right|$ limit, the approximation ${\tilde {m}}$ (\ref{eq:mtildesmallell}) becomes increasingly accurate.

We emphasize an important point here.
In our ${\mathfrak {su}}(3)$ numerical examples above, we fixed $r_{h}$, $\omega _{1}(r_{h})$ and $\omega _{2}(r_{h})$ and examined the
solutions for varying values of the cosmological constant $\Lambda $.
However, this is {\em {not}} the purpose of the approximation developed in this section.
The approximation derived here is for {\em {fixed}} (but very large) $\Lambda $, and varying $r_{h}$ (keeping $r_{h}\gg \ell $),
$\omega _{1}(r_{h})$ and $\omega _{2}(r_{h})$ (subject to $q_{j}$ (\ref{eq:qdef}) being of order one for small $\ell $).
We think of the approximation $\omega _{j}={\mbox {constant}}$, $\eta _{j}$ (\ref{eq:etasolve}), ${\tilde {m}}$ (\ref{eq:mtildesmallell})
as being the first term in an asymptotic series for the field variables, which is asymptotic for large $\left| \Lambda \right|$.
This approximation is uniformly valid on a region of the parameter space in which $r_{h}\gg \ell$, and $\sigma $ (\ref{eq:qdef}) is fixed between
$\frac {1}{3}$ and $1$.
This does not cover the whole space of black hole solutions for this large, but fixed, value of $\left| \Lambda \right| $.
However, this is not unexpected: the original existence theorems \cite{Winstanley1,Baxter3} were only proved for $\left| \Lambda \right| $
``sufficiently large'' and fixed $r_{h}$ and $\omega _{j}$.
It is therefore unsurprising that we are only able to derive analytic approximations on some subspace of the set of black hole solutions.
However, the space of black hole solutions that we are able to describe is large and increases in size as $\ell $ decreases.

Having derived the approximate solutions $\omega _{j}={\mbox {constant}}$, $\eta _{j}$ (\ref{eq:etasolve}),
${\tilde {m}}$ (\ref{eq:mtildesmallell})
for small $\ell $, we now examine whether these approximate solutions are determined by their mass $M$ and non-Abelian charges $Q_{i}$.
For these approximate solutions, the gauge field functions $\omega _{j}$ are
approximately constant, and therefore the charges (\ref{eq:Qjfinal}) are given,
 to leading order in $\ell $, in terms
of the values of the gauge field functions on the event horizon:
\begin{equation}
Q_{j} = \frac {{\sqrt {j(j+1)}}}{{\sqrt {2}}}
\left( 1 - \frac {\omega _{j}(r_{h})^{2}}{j} + \frac {\omega _{j+1}(r_{h})^{2}}{j+1}
\right) .
\end{equation}
In addition, the masses of the black holes are given, to leading order in $\ell $, by
\begin{equation}
M =
\frac {r_{h}}{2} - \frac {\Lambda r_{h}^{3}}{6}
+\frac {1}{4r_{h}}
\sum _{j=1}^{N}
\left[\left(\omega^2_j(r_{h})-\omega^2_{j-1}(r_{h})-N-1+2j\right)^2\right] .
\end{equation}
To show that the Jacobian $J$ (\ref{eq:Jacobian}) does not vanish in this case, we
require the partial derivatives
\begin{eqnarray}
\frac {\partial M}{\partial r_{h}} & = &
\frac {1}{2} - \frac {\Lambda r_{h}^{2}}{2}
- \frac {1}{4r_{h}^{2}} \sum _{j=1}^{N}
\left[\left(\omega^2_j(r_{h})-\omega^2_{j-1}(r_{h})-N-1+2j\right)^2\right] ;
\nonumber \\
\frac {\partial M}{\partial \omega _{k}(r_{h})} & = &
-\frac {2}{r_{h}} W_{k}(r_{h}) \omega _{k}(r_{h}) ;
\nonumber \\
\frac {\partial Q_{j}}{\partial r_{h}} & = &
0 ;
\nonumber \\
\frac {\partial Q_{j}}{\partial \omega _{k}(r_{h})} & = &
\frac {{\sqrt {j(j+1)}}}{{\sqrt {2}}}
\frac {2\omega _{k}(r_{h})}{k} \left(  - \delta _{j,k} + \delta _{j+1,k} \right) .
\end{eqnarray}
The fact that the $Q_{j}$ do not depend on $r_{h}$
means that the Jacobian $J$ (\ref{eq:Jacobian})
is given by
\begin{equation}
J = \frac {\partial M}{\partial r_{h}} J_{Q},
\label{eq:reducedJ}
\end{equation}
where $J_{Q}$ is the Jacobian of the charges $Q_{j}$ in terms of the
$\omega _{k}(r_{h})$.
We observe that $\frac {\partial M}{\partial r_{h}}>0$ follows immediately from
(\ref{eq:constraint}), while $J_{Q}$ must be non-zero by virtue of the fact that the
transformation (\ref{eq:Qjfinal}) from the values of the $\omega _{k}$ to the charges $Q_{j}$ is invertible.
Therefore the Jacobian $J$ (\ref{eq:reducedJ}) is non-zero.
This means that the parameters $\Lambda $, $M$ and $Q_{j}$
uniquely specify the black hole solutions,
at least in this approximation.

The approximation we have used in this section is exact in the limit
$\ell \rightarrow 0$.
We have therefore shown that the Jacobian $J$ is non-zero in this limit.
Using the fact that the Jacobian $J$ is analytic in the parameters of the theory,
as shown in section \ref{sec:analyticity}, it follows that $J\neq 0$ at least for
sufficiently small $\ell $.
We therefore deduce that
at least a subset of stable hairy
${\mathfrak {su}}(N)$ black holes, for sufficiently
large $\left| \Lambda \right| $, are uniquely characterized by their mass, the
(negative) cosmological constant $\Lambda $ and a set of global conserved non-Abelian
charges $Q_{j}$.

A further comment is in order.  In previous sections, we have emphasized that
Bizon's modified ``no-hair'' conjecture \cite{Bizon2} applies only to stable black holes.  In our numerical work in section \ref{sec:numerical}, we restricted our attention to ``potentially stable'' black holes, namely those for which the inequalities (\ref{eq:stabineqs}), necessary (but not sufficient) for stability, are satisfied.  However, in this section we have made no reference to these inequalities, which, with the definition (\ref{eq:qdef}), take the form
\begin{equation}
q_{j}(x) > \frac {1}{2} \left[ q_{j+1}(x) + q_{j-1} (x) \right].
\label{eq:qstabineqs}
\end{equation}
Simply by restricting our attention to those black holes for which (\ref{eq:qstabineqs}) hold at the event horizon $x=1$, we trivially have that stable black holes are uniquely specified by $\Lambda $, $M$ and the charges $Q_{j}$, for
sufficiently small $\ell $.

\subsection{Characterizing solitons by $M$ and $Q_{j}$}
\label{sec:solitons}

In the previous subsection we have shown that ${\mathfrak {su}}(N)$ black holes,
at least for sufficiently large black holes in the
presence of a sufficiently large $\left| \Lambda \right| $, are uniquely
characterized by their mass $M$, the ${\mathfrak {su}}(N)$ charges
$Q_{j}$ and $\Lambda $.
One remaining issue is whether it is possible for soliton solutions to have the
same $M$ and $Q_{j}$ as black hole solutions for a particular $\Lambda $.
In other words, is it possible to confuse black holes and solitons by measuring their
mass $M$ and non-Abelian charges $Q_{j}$?

Our analysis mirrors that in the previous subsection.
However, as in \cite{Baxter3}, we find that analytic work with soliton solutions is
considerably more complicated than that for black hole solutions.
The work in this section follows the approach and notation of \cite{Baxter3}
in dealing with the soliton solutions, and we refer the reader
to that paper for more details, keeping the presentation in this section brief.

Our goal is to find an approximation for the soliton solutions
which is valid for small $\ell $.  Since there is no event horizon for
soliton solutions, we have just one length scale, namely $\ell $, and we
define a new dimensionless radial variable $y$ by $y=r/\ell $.
Our first task is to write the field equations (\ref{eq:YMe}, \ref{eq:Ee})
in a form suitable for analysis.
Following \cite{Baxter3}, we write the gauge field functions $\omega _{j}(r)$ as
\begin{equation}
\omega _{j}(r) = \left[ j \left( N-j \right) \right] ^{\frac {1}{2}} u_{j}(y)
\label{eq:omegasol}
\end{equation}
and define a vector ${\bmath {u}} = \left( u_{1}, \ldots , u_{N-1} \right) ^{T}$.
As in \cite{Baxter3}, we rewrite the vector ${\bmath {u}}$ as a sum over
eigenvectors of the $\left( N-1\right) \times \left( N-1 \right)$
matrix $A$ whose entries are:
\begin{equation}
A_{i, j} = \left[ j \left( N - j\right) \right] ^{\frac {1}{2}}
\left[ 2\delta _{i,j} - \delta _{i+1,j} - \delta _{i-1,j} \right] ,
\label{eq:matrixA}
\end{equation}
where $\delta _{i,j}$ is the usual Kronecker delta.
The form of ${\bmath {u}}$ in terms of eigenvectors of $A$ reads
(cf. (77) in \cite{Baxter3}):
\begin{equation}
{\bmath {u}}(y) = {\bmath {u}}_{0} + \sum _{k=2}^{N}
{\bmath {\beta }}_{k}(y) y^{k} \ell ^{k}
\label{eq:usol}
\end{equation}
where ${\bmath {u}}_{0} = \left( 1, 1, \ldots, 1 \right) ^{T}$ and the
${\bmath {\beta }}_{k}$ are vector functions which
satisfy
\begin{equation}
A{\bmath {\beta }}_{k} = k \left( k -1 \right) {\bmath {\beta }}_{k}.
\end{equation}
Next we define scalar variables $\zeta _{k}(y)$ by (cf. (81) in \cite{Baxter3})
\begin{equation}
\zeta _{k}(y) = {\bmath {\sigma }}_{k}^{T}{\bmath {\beta }}_{k}(y) ,
\label{eq:zetasol}
\end{equation}
where ${\bmath {\sigma }}_{k}^{T}$ is the $k-$th left eigenvector of the matrix $A$.

We next define a rescaled metric variable ${\hat {m}}(y)$ as follows:
\begin{equation}
{\hat {m}}(y) = \frac {m(r)}{\ell },
\end{equation}
which satisfies the Einstein equation (\ref{eq:Ee}):
\begin{equation}
\frac {d{\hat {m}}}{dy} = \mu G + \frac {1}{2y}{\tilde {P}}
\label{eq:dmsol1}
\end{equation}
where $G$ is defined in (\ref{eq:Gdef}) and
\begin{equation}
{\tilde {P}}(y) = 2y^{3}\ell ^{2} p_{\theta },
\label{eq:tildePdef}
\end{equation}
with $p_{\theta }$ given by (\ref{eq:ptheta}).

It is shown in \cite{Baxter3} that the soliton solutions are determined,
in a neighbourhood of the origin, by the $N$ parameters
$\Lambda $ (or equivalently, $\ell $) and $\zeta _{k}(0)$\footnote{As
with the black hole solutions, strictly speaking there is an additional
parameter, namely $S(0)$, but we do not need to consider this further.}.
Unlike the black hole solutions, there are no a priori bounds on the
values of $\zeta _{k}(0)$ for the existence of regular solutions.
However, numerical analysis \cite{Baxter2} shows that the region of
parameter space for which soliton solutions in which all the gauge
field functions $\omega _{j}$ have no zeros expands as $\ell $ decreases.
It is argued in \cite{Breitenlohner1} that, for ${\mathfrak {su}}(2)$
solitons, the size of the parameter space expands as $\ell ^{-1}$
as $\ell $ decreases.
With this in mind, following (\ref{eq:qdef}), we define new variables
 $\alpha _{k}(y)$ by
\begin{equation}
\zeta _{k}(y) = \alpha _{k}(y) \ell ^{\sigma _{k}-1},
\label{eq:solzeta}
\end{equation}
where each $\sigma _{k}$ is a constant, and we expect that $\sigma _{k}<1$
so that the space of soliton solutions that we are considering expands as
$\ell $ decreases ($\sigma _{k} = 0$ would correspond to the results for
${\mathfrak {su}}(2)$ solitons \cite{Breitenlohner1}, but, as with
the black holes, we expect to not be able to consider the entire solution space,
but nonetheless a region of solutions which expands as
$\ell $ decreases).
Note that, unlike the black hole case (\ref{eq:qdef}), we allow for the
possibility of different values of $\sigma _{k}$ for
different $\zeta _{k}$.
We will assume that each $\alpha _{k}(y)$ is order one for all
$y\in \left[ 0, \infty \right)$ and small $\ell $.

The Yang-Mills equation (\ref{eq:YMe}) now becomes
\begin{eqnarray}
0  & = &  y^{2}\mu \left[ y^{k} \frac {d^{2}\alpha _{k}}{dy^{2}} + 2ky^{k-1}
\frac {d\alpha _{k}}{dy}
+ k \left( k -1 \right) y^{k-2}  \alpha _{k} \right]
\nonumber
\\ & &
+ \left[ 2{\hat {m}} + 2y^{3}-{\tilde {P}} \right] \left[ y^{k}
\frac {d\alpha _{k}}{dy} + ky^{k-1} \alpha _{k} \right] +
 \frac {1}{\ell ^{k+\sigma _{k}-1}}{\bmath {\sigma }}_{k}^{T} {\bmath {W}},
\label{eq:solYM_revised}
\end{eqnarray}
where ${\tilde {P}}$ is given by (\ref{eq:tildePdef}),
and the vector ${\bmath {W}}$ is defined as ${\bmath {W}} =
\left( W_{1}, W_{2}, \ldots , W_{N-1} \right) ^{T}$, with the $W_{j}$
given by (\ref{eq:Wdef}).

To determine the leading order (in $\ell $, for small $\ell $) behaviour
of the Yang-Mills equation (\ref{eq:solYM_revised}),
we first consider the
term ${\bmath {\sigma }}_{k}^{T} {\bmath {W}}$.
It is shown in \cite{Baxter3} that this takes the form
\begin{equation}
{\bmath {\sigma }}_{k}^{T} {\bmath {W}} = -k\left( k -1 \right)
y^{k}\ell ^{k+\sigma _{k}-1} \alpha _{k}
+ \sum _{j=k+1}^{Z} {\bmath {\sigma }}_{k}^{T} {\bmath {\tau }}_{j} y^{j} \ell ^{j},
\label{eq:sigmaTW}
\end{equation}
for some $Z\in {\mathbb {N}}$.
The ${\bmath {\sigma }}_{k}^{T} {\bmath {\tau }}_{j}$ are rather complicated
expressions which involve products of up to three
of the $\zeta _{k}$.
Since each $\sigma _{k}<1$, the leading order behaviour of
${\bmath {\sigma }}_{k}^{T} {\bmath {\tau }}_{j}$ will be from terms involving
products of three $\zeta _{k}$, which will be of order
$\ell ^{j-3+\sigma _{a}+\sigma _{b}+\sigma _{c}}$ for some $a$, $b$, $c$.
These will be subleading compared to the first term in (\ref{eq:sigmaTW})
if $k+\sigma _{k}-1<j-3+\sigma_{a}+\sigma _{b}+\sigma _{c}$,
bearing in mind that $j>k$.
This inequality is satisfied if $1>\sigma _{j}>\frac {2}{3}$ for all $j$.
Therefore, to find the leading order behaviour of the Yang-Mills equation
(\ref{eq:solYM_revised}), we need to keep only the first term in (\ref{eq:sigmaTW}).
This gives, to leading order in $\ell $,
\begin{eqnarray}
0  & = &  \mu \left[ y^{2} \frac {d^{2}\alpha _{k}}{dy^{2}} + 2ky
  \frac {d\alpha _{k}}{dy}
+ k \left( k -1 \right)  \alpha _{k} \right]
\nonumber
\\ & &
+ \left[ 2{\hat {m}}+ 2y^{3}-{\tilde {P}} \right] \left[
\frac {d\alpha _{k}}{dy} + ky^{-1} \alpha _{k} \right]
-k\left( k -1 \right) \alpha _{k} .
\label{eq:solYMrevised1}
\end{eqnarray}

Further simplification of the equations (\ref{eq:dmsol1}, \ref{eq:solYMrevised1})
requires analysis of the quantities $G$ and ${\tilde {P}}$.
First we write the vector functions ${\bmath {\beta }}_{k}$ in terms of the scalar
variables $\zeta _{k}$ (\ref{eq:zetasol}) (with no summation implied):
\begin{equation}
{\bmath {\beta }}_{k}(y) = {\bmath {v}}_{k} \zeta _{k}(y)
= {\bmath {v}}_{k} \ell ^{\sigma _{k} -1} \alpha _{k} (y),
\end{equation}
where the ${\bmath {v}}_{k} = \left( v_{k,1}, v_{k,2}, \ldots , v_{k,N-1} \right) $
are right-eigenvectors of the matrix $A$ (\ref{eq:matrixA}).
Using the expressions (\ref{eq:omegasol}, \ref{eq:usol}), we find
\begin{equation}
\omega _{j} = \left[ j \left( N - j \right) \right] ^{\frac {1}{2}}
\left[ 1 + \sum _{k=2}^{N} v_{k,j} y^{k}\ell ^{k+\sigma _{k} -1} \alpha _{k}(y)
\right]
\label{eq:omegasol1}
\end{equation}
and therefore the leading order behaviour of $G$ (\ref{eq:Gdef}) is
\begin{equation}
G = \ell ^{2\sigma _{2}} \Sigma _{G}
\left[ y^{2} \frac {d\alpha _{2}}{dy} + 2y \alpha _{2} \right] ^{2}
+ o \left( \ell ^{2} \right) ,
\end{equation}
where
\begin{equation}
\Sigma _{G} = \sum _{j=1}^{N-1} j\left( N - j \right)  v_{2,j}^{2} .
\end{equation}
From (\ref{eq:tildePdef}, \ref{eq:omegasol1}), it can be seen that
${\tilde {P}}$ is a complicated sum of terms involving
products of at least two $\zeta _{k}$, or, equivalently, at least two $\alpha _{k}$.
Products of the form $\alpha _{k_{1}}\alpha _{k_{2}}$ in ${\tilde {P}}$
are multiplied by
coefficients of order $\ell ^{k_{1}+k_{2}+\sigma _{k_{1}}+\sigma _{k_{2}}-4}$
and therefore the leading order term of this form occurs when $k_{1}=2=k_{2}$ and is
of order $\ell ^{2\sigma _{2}}$.
Products of the form $\alpha _{k_{1}}\alpha _{k_{2}}\alpha _{k_{3}}$ are multiplied by
coefficients of order $\ell ^{k_{1}+k_{2}+k_{3}+\sigma _{k_{1}}+\sigma _{k_{2}}
+\sigma _{k_{3}} -5}$ and are sub-leading compared with the $O(\ell ^{2\sigma _{2}})$
term.
Similarly, products of the form
$\alpha _{k_{1}}\alpha _{k_{2}}\alpha _{k_{3}}\alpha _{k_{4}}$
are also sub-leading, and we deduce that the leading order behaviour of
${\tilde {P}}$ is
\begin{equation}
{\tilde {P}} =2 \ell ^{2\sigma _{2}} \Sigma _{P} \alpha _{2}^{2}y^{3}
+ o \left( \ell \right) ,
\end{equation}
where
\begin{equation}
\Sigma _{P} = \sum _{j=1}^{N-1} \left[ j \left( N - j\right) v_{2,j}
-\left( j -1 \right) \left( N - j +1 \right) v_{2, j-1} \right] ^{2}.
\end{equation}

Substituting for $G$ and ${\tilde {P}}$ in the Einstein equation (\ref{eq:dmsol1}),
a consistent, non-trivial solution exists when
\begin{equation}
{\hat {m}} = \ell ^{2\sigma _{2}} \chi (y),
\end{equation}
with $\chi (y) $ satisfying, to leading order in $\ell $, the differential equation
\begin{equation}
\frac {d\chi }{dy}= \left( 1 + y^{2} \right) \Sigma _{G}
\left[ y^{2} \frac {d\alpha _{2}}{dy} + 2y \alpha _{2} \right] ^{2}
+  \Sigma _{P} \alpha _{2}^{2}y^{2} .
\label{eq:xieqn}
\end{equation}
The ${\hat {m}}$ and ${\tilde {P}}$ terms in (\ref{eq:solYMrevised1})
can therefore be ignored to leading order in $\ell $ provided $\sigma _{2}>0$,
giving simplified Yang-Mills equations
\begin{equation}
0 =
 y \left( 1 + y^{2} \right) \frac {d^{2}\alpha _{k}}{dy^{2}}
+ 2 \left[ k + \left( k +  1 \right) y^{2} \right]  \frac {d\alpha _{k}}{dy}
+k\left( k + 1 \right) y \alpha _{k}.
\label{eq:YMsolfinal}
\end{equation}
The Yang-Mills equations (\ref{eq:YMsolfinal}) were also derived in \cite{Baxter3}
in the limit $\ell \rightarrow 0$, but here we have used a more subtle approximation
(by including the $\sigma _{k}$ in (\ref{eq:solzeta})).
Equations (\ref{eq:YMsolfinal}) have the following solution regular at the origin
\cite{Baxter3}:
\begin{equation}
\alpha _{k}(y) = {}_{2}F_{1} \left( \frac {1}{2}\left[ k + 1\right],
\frac {k}{2} ;  k+\frac {1}{2} ; - y^{2} \right)
\alpha _{k,0} ,
\label{eq:hypergeom}
\end{equation}
where ${}_{2}F_{1}$ is a hypergeometric function and $\alpha _{k,0}=\alpha _{k}(0)$.
The function (\ref{eq:hypergeom}) has magnitude bounded by
$\left| \alpha _{k,0} \right| $ and, as $y \rightarrow \infty $,
it tends monotonically to zero as $O(y^{-2k+4})$ for $k>2$, and as $O(y^{-2})$ for
$k=2$.
Then (\ref{eq:xieqn}) can be integrated, to give $\chi (y)$, which
is then a bounded function satisfying the required boundary conditions:
\begin{equation}
\chi (y) = O(y^{3}), \quad y\rightarrow 0; \qquad
\chi (y) = \chi _{\infty }+ O(y^{-1}), \quad y \rightarrow \infty .
\end{equation}
We therefore have a consistent set of solutions to the field equations, valid when
$\ell \ll 1$ and providing $\frac {2}{3}<\sigma _{k}<1$.

Returning to the original variables $m(r)$ and $\omega _{j}(r)$, we have
$m(r) = O\left( \ell ^{2\sigma _{2}+1} \right) $ for all $r$,
so that the mass of these soliton solutions is also $O(\ell ^{2\sigma _{2}+1})$.
On the other hand, the charges (\ref{eq:Qjfinal}) can take on large values
because
\begin{equation}
\omega _{j} = \left[ j \left( N - j\right) \right] ^{\frac {1}{2}}
\left[ 1 + \sum _{k=2}^{N} v_{k,j} r^{k} \alpha _{k} \ell ^{\sigma _{k}-1} \right]
\end{equation}
and we have $\sigma _{k}<1$.
Therefore, for small $\ell $, the soliton solutions all have negligibly small mass
(this is confirmed by numerical calculations).
This is in contrast to the black hole solutions considered in the previous
subsection, which have non-negligible mass provided $r_{h} \gg \ell $.
We therefore conclude that the soliton solutions cannot be mistaken for stable black
hole solutions by measuring the mass and non-Abelian charges.

\section{Conclusions}
\label{sec:conc}

In this paper we have revisited stable furry black holes in ${\mathfrak {su}}(N)$ EYM with a negative cosmological
constant $\Lambda $, examining the consequences for the ``no-hair'' conjecture.
These black holes are ``furry'' because they have potentially unbounded amounts of stable gauge field hair, corresponding to
$N-1$ gauge degrees of freedom for ${\mathfrak {su}}(N)$.
Bizon's \cite{Bizon2} reformulation of the ``no-hair'' conjecture states that, within this fixed matter model, stable black holes should be
uniquely characterized by their mass and a set of global charges.
Our purpose in this paper has been to investigate whether this modified ``no-hair'' conjecture holds for ${\mathfrak {su}}(N)$ EYM.

The space of black hole solutions of ${\mathfrak {su}}(N)$ EYM in anti-de Sitter space is extraordinarily rich.
As well as purely magnetic, spherically symmetric, black hole solutions (the focus of our work in this paper), there are also dyonic spherically symmetric black holes
with both electric and magnetic charges \cite{Bjoraker};  static axisymmetric solitons \cite{Radu2} and black holes \cite{Radu3} and rotating black holes \cite{Radu4},
as well as a plethora of soliton solutions of each of these classes \cite{Radu4}.
Furthermore, there are black holes with non-spherical event horizon topology \cite{Radu1}.
These solutions are known only numerically, and, given the complexity of the field equations, deriving general analytic results for all these
cases is challenging.
For general $N$, it is known that
for sufficiently large $\left| \Lambda \right| $, there exist purely magnetic, spherically symmetric, soliton and black hole solutions
which are stable to spherically symmetric, linear perturbations \cite{Baxter1,Baxter3,Baxter4}.

We do not claim to have made a comprehensive study of all families in this zoo of solutions.
Instead we have focussed on purely magnetic, static, spherically symmetric black holes.
Within this simplified, restricted model, we have been able to present numerical and analytic evidence that black holes which are
stable both thermodynamically and under linear, spherically symmetric perturbations, are uniquely determined by their mass $M$ and a
set of $N-1$ non-Abelian global charges $Q_{j}$, constructed following \cite{Chrusciel}.
Our analytic argument is based on the first term of an asymptotic series for the metric and gauge field functions, valid in the limit
as the adS radius of curvature $\ell \rightarrow 0$.
Furthermore, this series is only applicable for a subspace of the full space of black hole solutions in this limit: in particular, we
restricted attention to large black holes with event horizon radius $r_{h}\gg \ell $ and a (albeit large) subset of the parameter space of values
of the gauge field functions on the event horizon.

With these limitations, we have provided evidence that Bizon's modified ``no-hair'' conjecture holds for at least this subclass of
furry black holes in ${\mathfrak {su}}(N)$ EYM with $\Lambda <0$.
Of course, it would be of great interest to extend these results to other families of black hole solutions within this matter model, particularly
dyonic black holes and rotating black holes.
However, we leave these problems for future work.

\ack
We thank Robert Bartnik for helpful discussions.
EW thanks the Universities of Monash, Calgary and the Australian National University
Canberra
for hospitality while this work was completed.
The work of BLS is supported by a studentship from EPSRC (UK).
The work of EW is supported by the Lancaster-Manchester-Sheffield
Consortium for Fundamental Physics under STFC grant ST/J000418/1, and by the European Co-operation in
Science and Technology (COST) action MP0905 ``Black Holes in a Violent Universe''.

\end{document}